\documentclass[journal,twoside,web]{IEEEtran}

\usepackage{cite}
\usepackage{amsmath,amssymb,amsfonts}
\usepackage{graphicx}
\usepackage{textcomp}
\usepackage{amsthm}

\theoremstyle{definition}

\newtheorem{theorem}{Theorem}
\newtheorem{lemma}{Lemma}
\newtheorem{remark}{Remark}
\newtheorem{assumption}{Assumption}

\usepackage{algorithm}
\usepackage{algorithmic}

\def\BibTeX{{\rm B\kern-.05em{\sc i\kern-.025em b}\kern-.08em
		T\kern-.1667em\lower.7ex\hbox{E}\kern-.125emX}}
\begin{document}
	\title{Distributed Zonotopic Fusion Estimation for Multi-sensor Systems}
	\author{Yuchen Zhang, Bo Chen*, \IEEEmembership{Senior Member,~IEEE}, Zheming Wang, \IEEEmembership{Member, IEEE}, \\ Wen-An Zhang,~\IEEEmembership{Senior Member,~IEEE}, Li Yu, \IEEEmembership{Senior Member,~IEEE}, and Lei Guo, \IEEEmembership{Fellow,~IEEE}
		\thanks{This work was supported in part by the National Natural Science Funds of China under Grant 61973277 and Grant 62073292, in part by the Zhejiang Provincial Natural Science Foundation of China under Grant LR20F030004, and in part by the Key Research and Development Program of Zhejiang Province under Grant 2023C01144. (* Corresponding author)}
		\thanks{Yuchen Zhang, Bo Chen, Zheming Wang, Wen-An Zhang, and Li Yu are with the Department of Automation, Zhejiang University of Technology, Hangzhou 310023, China (email: YuchenZhang95@163.com, bchen@aliyun.com, wangzheming@zjut.edu.cn, wazhang@zjut.edu.cn, lyu@zjut.edu.cn).}
		\thanks{Lei Guo is with the School of Automation Science and Electrical Engineering, Beihang University, Beijing 100191, China (email: lguo@buaa.edu.cn)}
		}
	
	\maketitle
	
	\begin{abstract}
		Fusion estimation is often used in multi-sensor systems to provide accurate state information which plays an important role in the design of efficient control and decision-making. This paper is concerned with the distributed zonotopic fusion estimation problem for multi-sensor systems. The objective is to propose a zonotopic fusion estimation approach using different zonotope fusion criteria. We begin by proposing a novel zonotope fusion criterion to compute a distributed zonotopic fusion estimate (DZFE). The DZFE is formulated as a zonotope enclosure for the intersection of local zonotopic estimates from individual sensors. Then, the optimal parameter matrices for tuning the DZFE are determined by the analytical solution of an optimization problem. To reduce the conservatism of the DZFE with optimal parameters, we enhance our approach with an improved zonotope fusion criterion, which further improves the estimation performance of this DZFE by constructing tight strips for the intersection. In addition, we tackle the problem of handling sequentially arrived local estimates in realistic communication environments with a sequential zonotope fusion criterion. This sequential zonotope fusion offers reduced computational complexity compared to batch zonotope fusion. Notice that the proposed zonotope fusion criteria are designed to meet the state inclusion property and demonstrate performance superiority over local zonotopic estimates. We also derive stability conditions for these DZFEs to ensure their generator matrices are ultimately bounded. Finally, two illustrative examples are employed to show the effectiveness and advantages of the proposed methods.
	\end{abstract}
	
	\begin{IEEEkeywords}
		Multi-sensor system, zonotope fusion criterion, sequential zonotope fusion, state inclusion property, performance superiority.
	\end{IEEEkeywords}
	
	\section{Introduction}
	\subsection{Background and Motivation}
	In recent decades, multi-sensor fusion estimation has attracted considerable research interest due to its broad applications in target tracking \cite{mahler2007statistical, battistelli2013consensus, fantacci2018robust}, localization \cite{zhang2016sequential, gao2020randomfinitesetbased}, environmental monitoring \cite{bai2018collaborative, chen2017distributed}, and health monitoring \cite{nathan2018particle, king2017application}. The primary objective of fusion estimation is to enhance state estimation by integrating valuable information from multiple sensors. For example, raw measurements from individual sensors can be transmitted to a fusion center (FC) for measurement fusion \cite{ran2009two, zhang2014twolevel, roecker1988comparison} or centralized fusion estimation \cite{ma2013centralized, chang1997optimal, li2003optimal}. Alternatively, the FC may aggregate local estimates from sensors for distributed fusion estimation \cite{chen2014distributeda, yan2013optimal, sun2017multisensor}. Compared to centralized fusion estimation, distributed fusion has better robustness, flexibility, and reliability due to its parallel structure \cite{sun2017multisensor}. In the event of a sensor fault, local fault detection and isolation mechanisms can prevent performance degradation in a distributed fusion structure.
	
	A key challenge in distributed fusion estimation is the design of effective fusion criteria to integrate local estimates from sensors and enhance estimation accuracy. Most existing fusion criteria are tailored for Gaussian posterior estimates. However, noise in practical applications may not strictly follow a Gaussian distribution, and obtaining its statistical information accurately is also challenging. As set-membership estimation provides sets that enclose the admissible values of the state, it is crucial to develop new fusion criteria for set-based estimates. In particular, the zonotopic estimation framework has advantages in reduced computational complexity and minimized wrapping effects caused by the propagation of overestimates \cite{kuhn1998rigorously}. Therefore, this paper addresses this need by focusing on the design of zonotope fusion criteria for distributed zonotopic fusion estimation.
	
	\subsection{Contributions}
	The following contributions are made in this paper:
	
	\vspace{4pt}
	\begin{itemize}
		\item[1)] \textbf{Zonotope Fusion Criteria:} We propose a zonotope fusion criterion for merging local zonotopic estimates into a distributed zonotopic fusion estimate (DZFE). The optimal parameter matrices for this criterion are obtained through the analytical solution of an optimization problem. Thereafter, we introduce an improved zonotope fusion criterion by constructing tight strips for the intersection of local zonotopic estimates to enhance the DZFE with optimal parameters. Compared to our previous work \cite{zhang2023distributed}, the new zonotope fusion criteria not only ensure the state inclusion property but also demonstrate performance superiority over local zonotopic estimates. In other words, DZFEs that integrate information from multiple sensors can provide better estimation performance than local estimates using a single sensor.
		
		\item[2)] \textbf{Sequential Zonotope Fusion:} To reduce the computational burden, we propose a sequential zonotope fusion criterion that fuses local zonotopic estimates sequentially. This approach distributes the computational cost of the sequential DZFE over the entire time period by splitting the high-dimensional batch fusion process into multiple low-dimensional fusion steps. Furthermore, the sequential DZFE maintains the state inclusion property and the performance superiority in the batch fusion criteria.
		
		\item[3)] \textbf{Stability Analysis:} We conduct a stability analysis of the proposed DZFEs based on their performance superiority. The results indicate that the DZFEs are ultimately bounded if at least one of the local zonotopic estimates is ultimately bounded. This implies that the performance degradation of a local estimate has a limited impact on the performance of the DZFEs.
	\end{itemize}
	
	\subsection{Related Works}
	It is widely recognized that the design of fusion criteria plays a central role in distributed fusion estimation for multi-sensor systems. According to the availability of local estimation error correlations, most existing fusion criteria can be classified into two categories: C1) fusion criteria with known correlations; C2) fusion criteria with unknown correlations. A detailed review of these fusion criteria is provided below.
	\begin{itemize}
		\item[C1)] When estimation error correlations are known, optimal fusion criteria can be designed by augmenting or weighting the local estimates. In \cite{kim1994development}, an optimal fusion criterion in the maximum likelihood sense was designed for Gaussian posterior estimates. Subsequently, unified optimal fusion criteria for general linear data models, including raw measurement data and local estimates, were proposed in the best linear unbiased estimation sense and the weighted least squares sense \cite{li2003optimal}. Inspired by the work in \cite{kim1994development}, optimal fusion criteria weighted by scalars \cite{sun2005distributed}, diagonal matrices \cite{sun2004multisensora}, and general matrices \cite{sun2004multisensor} were developed in the linear unbiased minimum variance sense. It was proven that the optimal fusion criterion weighted by general matrices \cite{sun2004multisensora} is equivalent to the maximum likelihood fusion criterion \cite{kim1994development} for Gaussian posterior estimates, and also equivalent to the weighted least squares fusion criterion \cite{li2003optimal}.
		\item[C2)] When estimation error correlations are unknown, the fusion criteria should yield consistent estimates regardless of the actual correlations. The well-known covariance intersection (CI) algorithm was proposed in \cite{julier1997nondivergent} by using the convex combination of local estimates' covariances. Thereafter, it was pointed out that the CI technique is equivalent to the log-linear combination of local estimates' Gaussian functions and can be generalized to a fusion criterion for any probability density functions \cite{mahler2000optimal, hurley2002information, julier2006using, clark2010robust}. More recently, other fusion criteria based on probabilistic density function fusion have been developed \cite{koliander2022fusion, beekman2020review}.
	\end{itemize}
	
	The fusion criteria discussed above are tailored for Gaussian posterior estimates or estimates with known probability distributions. However, it is not easy to obtain statistical information about noises and validate Gaussian assumptions in practical scenarios. In the absence of precise statistical information, local sensors often use set-membership estimators to recursively compute sets that enclose possible states. A variety of set-membership estimators have been developed based on the properties of different set representations. For example, intervals are used to design interval observers \cite{gouze2000interval, mazenc2011interval, raissi2012interval, combastel2013stable, efimov2013interval} due to their computational efficiency in numerical implementation. Ellipsoids also offer low-cost computations and are used for developing different ellipsoidal estimators \cite{a.1997ellipsoidal, polyak2004ellipsoidal, durieu2001multiinput, bertsekas1971recursive, schweppe1968recursive}. General polytopes can describe set domains with any shape but encounter complexity issues when enumerating edges and vertices in high-dimensional spaces {\cite{ziegler2012lectures}}. Actually, there is a trade-off between computation complexity and representation precision for different sets. Only certain special polytopes, such as parallelotopes \cite{chisci1996recursive, vicino1996sequential} and zonotopes \cite{puig2001worstcase, combastel2003state, alamo2005guaranteed, le2013zonotopic, combastel2015zonotopes, depaula2022zonotopic}, were used for designing set-membership estimators because they effectively balance these factors. Particularly, zonotopes can represent more sophisticated set domains than intervals, ellipsoids, and parallelotopes. Additionally, zonotopes can be fully characterized by a vector and a rectangular matrix, which simplifies zonotope operations to matrix calculations. Consequently, zonotopic estimation has gained increasing attention in recent research \cite{kousik2023ellipsotopes, scott2016constrained, combastel2020distributed, kochdumper2021sparse, combastel2022functional, althoff2021set}. This motivates the investigation of set-membership fusion estimation, especially zonotopic fusion estimation, to develop a new framework for multi-sensor systems. To date, research on fusion estimation under unknown but bounded noises is still in its early stages. A distributed fusion estimator under bounded noises with unknown lower and upper bounds was proposed in \cite{chen2019new}. Meanwhile, the ellipsoidal fusion estimation problem was addressed in \cite{wang2019ellipsoidal} using convex optimization, and zonotopic fusion estimators were developed by minimal parallelotope enclosure \cite{zhang2023distributed} and matrix-weighted fusion \cite{zhao2023zonotopic}. How to design fusion criteria to reduce the conservatism of fusion estimators remains an open question.
		
	\textit{Notation}: Throughout this paper, we use the following notations. The identity matrix is denoted by $\mathbf{I}$, and a column vector with all ones is represented as $\mathbf{1}$. Let us define $\mathbb{N}_L := \{1,2,...,L\}$, where $L$ is a natural number excluding zero. Given sets $\Sigma_1$ and $\Sigma_2$, the notation $\Sigma_1 \setminus \Sigma_2$ denotes the set of elements in $\Sigma_1$ that are not in $\Sigma_2$. The trace, transpose, and inverse of a matrix $A$ are represented by $\mathrm{Tr}(A)$, $A^{\mathrm{T}}$, and $A^{-1}$, respectively. The matrix $A$ is said to be `spd' if it is symmetric ($A=A^{\mathrm{T}}$) and positive definite ($A \succ 0$, i.e., $\forall x \neq 0$, $x^{\mathrm{T}}Ax > 0$). Moreover, the infinite norm of a vector $x$ is denoted by $\|x\|_{\infty}$, and the weighted Frobenius norm of a matrix $R \in \mathbb{R}^{n \times r}$ is defined as $\|R\|_{W} := \sqrt{\mathrm{Tr}(R^{\mathrm{T}}WR)}$, where $W \in \mathbb{R}^{n \times n}$ is a positive definite matrix. The notation $|\cdot|$ represents the element-by-element absolute value operator, and $\mathrm{diag}(x)$ is a diagonal matrix with diagonal elements from the vector $x$. Given a matrix $R \in \mathbb{R}^{n \times r}$, the submatrix consisting of columns $a$ through $b$ is denoted as $R_{:,a:b}$ (with $R_{:,a}$ representing the $a$th column), and $\vec{R}$ is the matrix obtained by sorting the columns of $R$ in decreasing order of their weighted vector norm, such that $\left\|\vec{R}_{:, i}\right\|_W^2 \ge \left\|\vec{R}_{:, i+1}\right\|_W^2$.
	
	\section{Preliminaries}
	In this section, we will first introduce four representations of a zonotope. The general-representation of an $r$-order zonotope $\mathcal{Z} \subset \mathbb{R}^n$ is given by:
	\begin{equation}
		\label{E8-1}
		\mathcal{Z} = \langle c, R \rangle := \{z \in \mathbb{R}^n: z=c+Ru, \ \|u\|_{\infty} \le 1 \}
	\end{equation}
	where $c \in \mathbb{R}^n$ and $R \in \mathbb{R}^{n \times r}$ are the center and the generator matrix of $\mathcal{Z}$, respectively. Actually, the zonotope $\mathcal{Z}$ is an affine transformation of the unit hypercube $\mathcal{B}^r := \{u \in \mathbb{R}^r: \|u\|_{\infty} \le 1\} \subset \mathbb{R}^r$. As a special kind of polytope, zonotopes also have $\mathcal{H}$-representations and $\mathcal{V}$-representations \cite{ziegler2012lectures}. The $\mathcal{H}$-representation describes the zonotope $\mathcal{Z}$ as the intersection of closed half-spaces:
	\begin{equation}
		\label{E8-2}
		\mathcal{Z}=\{ z \in \mathbb{R}^n: Hz \le b \}
	\end{equation}
	where $H = \mathrm{col} \{ h_1, -h_1, \cdots, h_{\varepsilon}, -h_{\varepsilon} \} \in \mathbb{R}^{2\varepsilon \times n}$, $b = \mathrm{col} \{ b_1, \cdots, b_{2\varepsilon} \} \in \mathbb{R}^{2\varepsilon}$, and $2\varepsilon$ is the number of half-spaces. The $\mathcal{V}$-representation expresses the zonotope $\mathcal{Z}$ as the convex hull of a finite set of vertices:
	\begin{equation}
		\label{E8-3}
		\mathcal{Z} = \mathrm{cone}(V):= \left\{z \in \mathbb{R}^n: z= \sum\nolimits_{i=1}^{\kappa} \pi_i \nu_i, \ \pi_i \ge 0 \right\}
	\end{equation}
	where $\nu_i \in \mathbb{R}^{n \times 1}$ is a vertex of the zonotope $\mathcal{Z}$, $\kappa$ is the number of vertices, and $V = \left[ \nu_1,...,\nu_i, ..., \nu_{\kappa} \right] \in \mathbb{R}^{n \times \kappa}$. Moreover, the strip-representation of the zonotope $\mathcal{Z}$ can be obtained from (\ref{E8-2}) as the intersection of $\varepsilon$ tight strips:
	\begin{equation}
		\label{E8-4}
		\mathcal{Z} = \bigcap\nolimits_{i=1}^{\varepsilon} \mathcal{S}(c_i,d_i)
	\end{equation}
	where $\mathcal{S}(c_i,d_i):=\{z \in \mathbb{R}^n:|c_i z - d_i| \le 1\}$ denotes a strip, a region enclosed by two parallel hyperplanes. Its parameters are calculated as $c_i=\frac{2h_i}{b_{2i-1}+b_{2i}}$ and $d_i=\frac{b_{2i-1}-b_{2i}}{b_{2i-1}+b_{2i}}$. 
	
	Then, we will introduce the following operations for zonotopes. The Minkowski sum $\oplus$ of two zonotopes and the linear image $\odot$ of a zonotope by a matrix $L$ can be computed as follows \cite{combastel2015zonotopes}:
	\begin{subequations}
		\label{E8-5}
		\begin{align}
			& \langle c_1, R_1 \rangle \oplus \langle c_2, R_2 \rangle = \langle c_1+c_2, [R_1 \ \ R_2] \rangle  \\
			& L \odot \langle c, R \rangle = \langle Lc, LR \rangle.
		\end{align}
	\end{subequations}
	The order of the zonotope $\mathcal{Z}$ will increase rapidly with consecutive Minkowski sum operations, leading to higher demands on storage and computational resources. This issue can be mitigated using a weighted reduction operation such that:
	\begin{equation}
		\label{}
		\left\langle c,R \right\rangle  \subset  \left\langle c,\downarrow_{q,W}\!\!(R) \right\rangle 
	\end{equation}
	where $\downarrow_{q,W}(\cdot): \mathbb{R}^{n \times r} \rightarrow \mathbb{R}^{n \times q}$ is a weighted reduction operator \cite{combastel2015zonotopes} defined as
	\begin{equation}
		\label{E8-6}
		\downarrow_{q,W}\!\!(R) : = \left[ \vec{R}_{:,1:q-n}, \ b \left( \vec{R}_{:,q-n+1:r} \right)  \right].
	\end{equation}
	Here, $W$ is a given spd matrix, and $q \ (n \le q \le r)$ specifies the number of columns in the new generator matrix after the weighted reduction operation. This operator reorders the columns of $R$ in decreasing order of their weighted vector norm as a new matrix $\vec{R}$, and then encloses the submatrix $\vec{R}_{:,q-n+1:r}$, containing the less important columns of $R$, into an aligned box $b(\vec{R}_{:,q-n+1:r})$. The aligned box of a matrix $R \in \mathbb{R}^{n \times r}$ is defined as
	\begin{equation}
		\label{E8-7}
		b(R):=
		\begin{cases}
			\mathrm{diag}(|R| \cdot \mathbf{1})  \ \ \ \ & \text{if} \ r>n
			\\
			R \ \ \ \ & \text{otherwise}.
		\end{cases}
	\end{equation}
	
	\begin{remark}
		\label{R8-1}
		Zonotope is an efficient set representation for set-membership estimation, and its advantages can be summarized as follows:
		\begin{itemize}
			\item[a)] Zonotopes can represent sets more accurately than ellipsoids when their generator matrices are sufficiently large.
			\item[b)] Zonotopes are more computational efficient than general polytopes because their operations in (\ref{E8-5}) involve only simple matrix calculations.
			\item[c)] The wrapping effect \cite{kuhn1998rigorously} in the zonotope propagation can be effectively controlled as the operations of Minkowski sum and linear image are closed for zonotopes.
		\end{itemize}
	\end{remark}
	
	The following lemma shows the convexity of the function of weighted Frobenius norm square.
	
	\begin{lemma}
		\label{L8-1}
		Given a spd matrix $W \in \mathbb{R}^{n \times n}$, the function of weighted Frobenius norm square $f(\cdot)=\|\cdot\|_W^2$ with $\mathbf{dom} \ f = \mathbb{R}^{n \times m}$ is convex.
	\end{lemma}
	
	\noindent \textbf{Proof.}
		Let $X,Y \in \mathbb{R}^{n \times m}$ and $0 \le \theta \le 1$. It is known from the definition of weighted Frobenius norm that
		\begin{equation}
			\label{E8-8}
			\begin{aligned}
				\|X\|_W^2 
				& = \mathrm{Tr}\left\{X^{\mathrm{T}}WX\right\} = \mathrm{Tr}\left\{XX^{\mathrm{T}}W\right\} 
				\\
				& = \mathrm{Tr}\left\{\sum_{i=1}^m \left( X_{:,i}X_{:,i}^{\mathrm{T}}W \right) \right\}
				\\
				& = \sum_{i=1}^m X_{:,i}^{\mathrm{T}}WX_{:,i}
			\end{aligned}
		\end{equation}
		and the convexity of $f(\cdot)$ follows by
		\begin{equation}
			\label{E8-9}
			\begin{aligned}
				& f(\theta X+(1-\theta)Y) - \theta f(X) - (1-\theta) f(Y)
				\\
				& = - \theta (1-\theta) \sum_{i=1}^m \left( (X_{:,i}-Y_{:,i})^{\mathrm{T}}W(X_{:,i}-Y_{:,i}) \right) < 0.
			\end{aligned}
		\end{equation}
		\qed
		\vspace{8pt}
	
	\section{Problem Formulation}
\begin{figure}[]
		\begin{center}
			\includegraphics[height=5.5cm, width=\columnwidth]{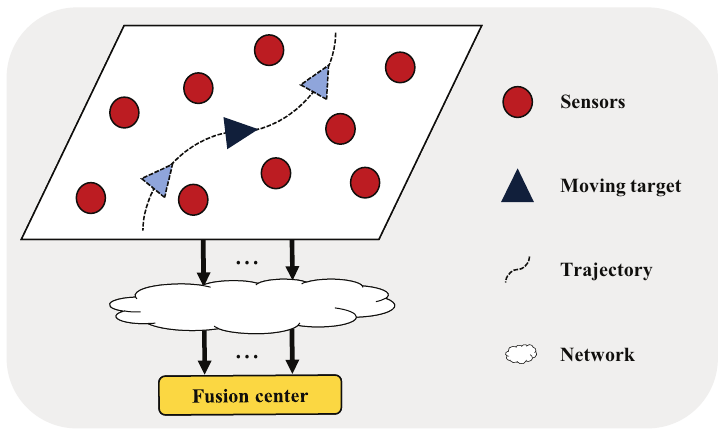}
			\caption{A multi-sensor system for moving target tracking.}
			\label{fig8-1}
		\end{center}
	\end{figure}

	In this section, we will formulate the problem of distributed zonotopic fusion estimation for multi-sensor systems. Consider a dynamic system monitored by $L$ sensors (see Fig. \ref{fig8-1} for a multi-sensor tracking system), where the state and measurement dynamics are described as follows:
	\begin{subequations}
		\label{E8-10}
		\begin{align}
			\label{E8-10.a}
			& x(k+1) = A(k)x(k) +	B(k)w(k) \\
			\label{E8-10.b}
			&	{y}_i(k) = C_i(k) x(k) + D_i(k) v_i(k) \ \ \ \ (i \in \mathbb{N}_L).
		\end{align}
	\end{subequations}
	Here, $x(k) \in \mathbb{R}^n$ represents the system state, and $y_i(k) \in \mathbb{R}^{m_i}$ denotes the measurement from the $i$th sensor. The bounded matrices $A(k)$, $B(k)$, $C_i(k)$, and $D_i(k)$ have appropriate dimensions. The system noise $w(k)$, the measurement noise $v_i(k)$, and the initial state $x(0)$ are assumed to be wrapped within the following zonotopic sets:
	\begin{equation}
		\label{E8-11}
		\begin{cases}
			& \!\!\!\!\!\!
			w(k) \in  {\mathbf{W}} := \langle \mathbf{0},{\mathbf{I}}_{{n}_{w}} \rangle  
			\\
			& \!\!\!\!\!\!
			{v}_i(k) \in {\mathbf{V}}_i := \langle \mathbf{0},{\mathbf{I}}_{{n}_{v_i}} \rangle  
			\\
			& \!\!\!\!\!\!
			{x}(0) \in  \langle {c}_{0},{R}_{0}\rangle \subset {\mathbb{R}}^{n}.
		\end{cases}
	\end{equation}
	
	A local zonotopic estimate can be recursively designed similar to a Kalman filter through three steps \cite{zhang2023distributed}: prediction, observation update, and reduction.
	\begin{itemize}
		\item[] \textbf{Prediction}: Based on the state dynamics (\ref{E8-10.a}), the prediction of $x(k)$ is as follows.
		\begin{equation}
			\label{E8-52}
			\begin{cases}
				& \!\!\!\!\!\!
				\hat{\mathcal{X}}_i^p(k) := \langle \hat x_i^p(k) , R_i^p(k) \rangle
				\\
				& \!\!\!\!\!\!
				\hat x_i^p(k) = A(k-1) \hat x_i(k-1)
				\\
				& \!\!\!\!\!\!
				R_i^p(k) = \big[A(k-1) R_i(k-1) \ \ B(k-1) \big].
			\end{cases}
		\end{equation}
		
		\item[] \textbf{Observation Update}: According to the measurement dynamics (\ref{E8-10.b}), the observation update is designed using the current measurement $y_i(k)$.
		\begin{eqnarray}
			\label{E8-53}
			\begin{cases}
				& \!\!\!\!\!\!
				\hat{\mathcal{X}}_i^o(k) := \langle \hat x_i^o(k) , R_i^o(k) \rangle
				\\
				& \!\!\!\!\!\!
				\hat x_i^o(k) = \hat x_i^p(k) + K_i(k) \left[ y_i(k) - C_i(k) \hat x_i^p(k) \right]
				\\
				& \!\!\!\!\!\!
				R_i^o(k) \!=\! \big[(\mathbf{I} \!-\! {K}_i(k){C}_i(k)) R_i^p(k) \ -\! K_i(k) D_i(k) \big].
			\end{cases}
		\end{eqnarray}
		
		\item[] \textbf{Reduction}: To reduce computational complexity, the weighted reduction operator (\ref{E8-6}) is employed to keep the number of columns in the generator matrix fixed at $r \ge n$:
		\begin{equation}
			\label{E8-54}
			\begin{cases}
				& \!\!\!\!\!\!
				\hat{\mathcal{X}}_i(k) := \langle \hat x_i(k) , R_i(k) \rangle
				\\
				& \!\!\!\!\!\!
				\hat x_i(k) = \hat x_i^o(k-1)
				\\
				& \!\!\!\!\!\!
				R_i(k) = \downarrow_{r,W}\!\!( R_i^o(k) ).
			\end{cases}
		\end{equation}
	\end{itemize}
	The reduction step is crucial as the local zonotopic estimator is operated on sensors with limited computational resources. Moreover, the optimal gain matrix $K_i^{\mathrm{opt}}(k)$, used throughout this paper, is determined as in \cite{zhang2023distributed, combastel2015zonotopes} by
	\begin{eqnarray}
		\label{E8-55}
		\begin{aligned}
			K_i^{\mathrm{opt}}(k) = 
			& R_i^p(k) \left[ R_i^p(k) \right]^{\mathrm{T}} C_i^{\mathrm{T}}(k)
			\\
			& \!\!\!\!\!\!\!\!\!\!\!\!\!\!\!\!\!\!\! \times \left\{ C_i(k)R_i^p(k) \left[ R_i^p(k) \right]^{\mathrm{T}} C_i^{\mathrm{T}}(k) \!+\! D_i(k) D_i^{\mathrm{T}}(k) \right\}^{-1}.
		\end{aligned}
	\end{eqnarray}
	It can also be proven that the state inclusion property $x(k) \in \hat{\mathcal{X}}_i(k)$ holds for all time instants if the initial estimate satisfies $x(0) \in \hat{\mathcal{X}}_i(0)$.
	
	Then, local zonotopic estimates $\hat{\mathcal{X}}_i(k) \ (i \in \mathbb{N}_L)$ are transmitted to a fusion center via a communication network. In the fusion center, a zonotope fusion step combines these estimates to yield a DZFE $\hat{\mathcal{X}}(k) := \langle \hat x(k), R(k)\rangle$ as follows:
	\begin{equation}
		\label{E8-13}
		\hat{\mathcal{X}}(k) = \mathcal{F}\left( \hat{\mathcal{X}}_{1}(k), ..., \hat{\mathcal{X}}_{i}(k), ..., \hat{\mathcal{X}}_{L}(k) \right).
	\end{equation}
	Here, the function $\mathcal{F}(\cdot)$ represents a zonotope fusion criterion to be designed. The estimation performance of a zonotopic estimate $\hat{\mathcal{X}}:=\langle c,R\rangle$ is evaluated based on the size of its wrapped region, which is fully characterized by its generator matrix. Therefore, the following performance index $J(\cdot)$ is defined and used in this paper to assess the estimation accuracy:
	\begin{equation}
		\label{E8-14}
		J\left( \hat{\mathcal{X}} \right)=J(R):= \| R \|_{W}^2.
	\end{equation}
	Generally, the DZFE (\ref{E8-13}) is expected to outperform local zonotopic estimates in terms of estimation accuracy, i.e., $J\left( \hat{\mathcal{X}}(k) \right) \le J \left( \hat{\mathcal{X}}_{i}(k) \right) \ (i \in \mathbb{N}_L)$. Additionally, the DZFE (\ref{E8-13}) must satisfy the state inclusion property, i.e., $x(k) \in \hat{\mathcal{X}}(k)$, and its computational complexity should be manageable for real-time implementation. Consequently, the aims of this paper can be summarized as follows.
	\begin{itemize}
		\item[1)] Design zonotope fusion criteria to ensure that the DZFEs provide better estimation performance than the local zonotopic estimates and maintain the state inclusion property.
		\item[2)] Design sequential zonotope fusion criteria to reduce the computational complexity of DZFEs.
		\item[3)] Derive stability conditions for the zonotopic fusion estimators to ensure that the estimation performance index of DZFEs is ultimately bounded.
	\end{itemize}
	
	\begin{remark}
		\label{R8-2}
		It should be emphasized that the zonotope fusion technique proposed in this paper is not limited to the distributed fusion estimation problem. For example, an estimator for linear time-invariant systems was proposed in \cite{xu2024observer} by minimizing the minimal robust positively invariant set, which can be extended to multi-sensor systems using the proposed zonotope fusion technique. Similarly, this technique can be applied to nonlinear systems provided that local nonlinear estimators \cite{alamo2005guaranteed, depaula2022zonotopic} are designed.
	\end{remark}
	
	\begin{remark}
		\label{R8-3}
		Notice that the local zonotopic estimates satisfy state inclusion property, meaning the system's real state is within the intersection of local zonotopic estimates. Although zonotopes are closed under the operations in (\ref{E8-5}), they are not closed under the intersection operation \cite{althoff2021set}. Therefore, the zonotope fusion criterion $\mathcal{F}(\cdot)$ aims to find a zonotope enclosure for the intersection of local zonotopic estimates. Similar to the fusion criteria for Gaussian posterior estimates, a performance index $J(\cdot)$ is necessary for evaluating and optimizing the fusion results, but here the zonotope fusion criteria utilize the size of zonotopes' wrapped region rather than the covariance matrix.
	\end{remark}
	
	\begin{remark}
		\label{R8-4}
		In \cite{althoff2011zonotope}, the concept `zonotope bundle' was introduced to describe the intersection of zonotopes and to facilitate the computation of reachable sets. However, the zonotope bundle propagates zonotopes separately without explicitly computing the intersection, leading to significant storage requirements and operational difficulties when new zonotopes are consecutively added to the zonotope bundle. This paper proposes using a zonotope enclosure of the intersection of original zonotopes for zonotope fusion, which is more computationally efficient and suitable for reachability analysis. This method also allows for the fusion of zonotopes in a zonotope bundle at the final step.
	\end{remark}
	
	\section{Main Results}
	In this section, we will present the zonotope fusion criteria and sequential zonotope fusion criteria, and then derive stability conditions for the proposed DZFEs.
	
	\subsection{Zonotope fusion criteria}
	The technique for finding the intersection between zonotopes and strips is detailed in \cite{alamo2005guaranteed}. Building on this work, we derive the following theorem as the first zonotope fusion criterion.
	
	\begin{theorem}
		\label{T8-1}
		Given local zonotopic estimates $\hat{\mathcal{X}}_{i}(k) \ (i=1,2,...L)$, a DZFE $\hat{\mathcal{X}}_{[1]}(k)$ can be determined by the following fusion criterion:
		\begin{eqnarray}
			\label{E8-15}
			\begin{cases}
				& \!\!\!\!\!\!
				\hat x_{[1]}(k) = \hat x_{1}(k) + \sum_{i=2}^L M_i(k) (\hat x_{i}(k)- \hat x_{1}(k))
				\\
				& \!\!\!\!\!\!\!
				\begin{aligned}
					R_{[1]}(k) \!=\!
					& \bigg[ \!\! \left( \mathbf{I} \!-\!\sum\nolimits_{i=2}^L \!M_i(k) \right)\!R_{1}(k) \ \ M_2(k)R_{2}(k) \\ 
					& \ \ \ \ \ \ \ \ \ \ \ \ \ \ \ \ \ \ \ \ \ \ \ \ \ \ \cdots \ \ \ M_L(k)R_{L}(k)\bigg]
				\end{aligned}
			\end{cases}
		\end{eqnarray}
		where $M_i(k) \in \mathbb{R}^{n \times n}$ $(i=2,...,L)$ are designed parameter matrices. Moreover, the DZFE obtained from (\ref{E8-15}) satisfies $x(k) \in \mathcal{O}(k) \subseteq \hat{\mathcal{X}}_{[1]}(k)$, where $\mathcal{O}(k):=\bigcap_{i \in \mathbb{N}_L} \hat{\mathcal{X}}_{i}(k)$ is the intersection of the local zonotopic estimates.
	\end{theorem}
	
	\noindent \textbf{Proof.}
		It is known that $x(k) \in \hat{\mathcal{X}}_{i}(k) \ (i \in \mathbb{N}_L)$, which implies $x(k) \in \mathcal{O}(k)$. According to the general-representation of a zonotope, there exists $u_i(k) \in \mathcal{B}^{r}$ such that $x(k) = \hat x_{i}(k) + R_{i}(k) u_i(k) \ (i=1,2,..,L)$. Equivalently, the following $L$ equations must all be satisfied $\forall x(k) \in \mathcal{O}(k)$:
		\begin{subequations}
			\label{E8-16}
			\!
			\begin{align}
				&
				x(k) - \hat x_{1}(k) - R_{1}(k) u_1(k) = 0
				\\
				&
				\hat x_{1}\!(k) + R_{1}(k) u_1(k) - \hat x_{i}(k) - R_{i}(k)  u_i(k)  = 0
			\end{align}
		\end{subequations}
		where $i = 2,3,...,L$. Note that (\ref{E8-16}) is a sufficient condition for the following equation:
		\begin{equation}
			\label{E8-17}
			x(k) = \hat x_{[1]}(k) + R_{[1]}(k) u(k)
		\end{equation}
		where $u(k):=\mathrm{col}\{u_1(k), ..., u_L(k)\} \in \mathcal{B}^{rL}$. Again, using the general-representation of a zonotope, the equation (\ref{E8-17}) implies that $x(k) \in \hat{\mathcal{X}}_{[1]}(k)$. Hence, $\mathcal{O}(k) \subseteq \hat{\mathcal{X}}_{[1]}(k)$ can be derived from (\ref{E8-16}) to (\ref{E8-17}).
		\qed
		\vspace{8pt}
	
	\begin{remark}
		\label{R8-5} Although a zonotope can be represented as the intersection of strips, it is conservative to continuously perform the algorithm in \cite{alamo2005guaranteed} to obtain the intersection of zonotopes. In contrast, Theorem 1 fuses all zonotopes simultaneously to avoid potential conservatism. Actually, the parameter matrices $M_i(k) \ (i=2,...,L)$ involve a total of $(L-1)n^2$ variables, more than the $(L-1)n$ tunable variables used by the technique in \cite{alamo2005guaranteed}.
	\end{remark}
	
	Let us augment all parameter matrices into $M(k):=\mathrm{row}\{M_2(k), ..., M_L(k)\}$. Then, the generator matrix in (\ref{E8-15}) can be rewritten as
	\begin{equation}
		\label{E8-18}
		R_{[1]}(k)=N_1(k) + M(k) N_2(k)
	\end{equation}
	where the matrices $N_1(k)$ and $N_2(k)$ are defined by
	\begin{subequations}
		\label{E8-19}
		\!
		\begin{align}
			& N_1(k) = [R_{1}(k) \ \ \mathbf{0} \ \ \cdots \ \ \mathbf{0}]
			\\
			& N_2(k) \!=\!
			\begin{bmatrix} 
				-R_{1}(k) & R_{2}(k) & \mathbf{0} & \cdots & \mathbf{0} \\
				-R_{1}(k) & \mathbf{0} & R_{3}(k) & \cdots & \mathbf{0} \\
				\vdots      & \vdots     & \vdots     & \ddots & \vdots     \\
				-R_{1}(k) & \mathbf{0} & \mathbf{0} & \cdots & R_{L}(k)
			\end{bmatrix}.
		\end{align}
	\end{subequations}
	To optimize the estimation performance of the designed DZFE, the following optimization problem with an analytical solution is formulated based on the performance index defined in (\ref{E8-14}).
	
	\begin{lemma}
		\label{L8-2}
		The optimal parameter matrices $M_i^{\mathrm{opt}}(k) \ (i=1,2...,L)$, which enable the designed DZFE (\ref{E8-15}) to achieve optimal estimation performance in (\ref{E8-14}), can be determined by solving the following optimization problem:
		\begin{equation}
			\label{E8-20}
			\begin{aligned}
				M^{\mathrm{opt}}(k) 
				& = \arg \min_{M(k)} J \left( R_{[1]}(k) \right)
				\\
				& = \arg \min_{M(k)} \left\| R_{[1]}(k) \right\|_W^2.
			\end{aligned}
		\end{equation}
		Moreover, the analytical solution to this optimization problem is given by
		\begin{equation}
			\label{E8-21}
			M^{\mathrm{opt}}(k) = -N_1(k) N_2^{\mathrm{T}}(k) \left( N_2(k) N_2^{\mathrm{T}}(k) \right)^{-1}.
		\end{equation}
	\end{lemma}
	
	\noindent \textbf{Proof.}
		By applying the definition of weighted Frobenius norm and (\ref{E8-18}), the derivative of the optimization objective (\ref{E8-20}) with respect to $M(k)$ is as follows:
		\begin{equation}
			\label{E8-22}
			\!
			\begin{aligned}
				\frac{\partial \left\| R_{[1]}(k) \right\|_W^2}{\partial M(k)} 
				& = \frac{\partial \mathrm{Tr}\left\{ N_2^{\mathrm{T}}(k) M^{\mathrm{T}}(k) W  M(k) N_2(k) \right\}}{\partial M(k)} 
				\\
				& \ \ \ \ + \frac{\partial \mathrm{Tr}\left\{ N_1^{\mathrm{T}}(k) W M(k) N_2(k) \right\}}{\partial M(k)} 
				\\
				& \ \ \ \ + \frac{\partial \mathrm{Tr}\left\{ N_2^{\mathrm{T}}(k) M^{\mathrm{T}}(k) W  N_1(k) \right\}}{\partial M(k)}.
			\end{aligned}
		\end{equation}
		Then using the properties of trace derivatives, one has that
		\begin{equation}
			\label{E8-23}
			\begin{aligned}
				\frac{\partial \left\| R_{[1]}(k) \right\|_{\!W}^2}{\partial M(k)} = 
				& 2W  M(k) N_2(k) N_2^{\mathrm{T}}(k) 
				\\
				& + 2W  N_1(k) N_2^{\mathrm{T}}(k).
			\end{aligned}
		\end{equation}
		It is known from Lemma \ref{L8-1} and (\ref{E8-18}) that $\left\| R_{[1]}(k) \right\|_W^2$ is convex with respect to $M(k)$. Therefore, the minimum value of $\left\| R_{[1]}(k) \right\|_W^2$ is achieved when $\partial \left\| R_{[1]}(k) \right\|_W^2 / \partial M(k) = 0$. By the weighted reduction operation in (\ref{E8-54}), $R_{i}(k)$ is known to have full row rank. Thus, $N_2(k)$ also has full row rank, making $N_2(k)N_2^{\mathrm{T}}(k)$ invertible. The analytical solution (\ref{E8-21}) follows from setting $\partial \left\| R_{[1]}(k) \right\|_W^2 / \partial M(k) = 0$.
		\qed
		\vspace{8pt}
	
	\begin{remark}
		\label{R8-6}
		While the volume of a zonotopic estimate can also be used to evaluate the size of its wrapped region, computing this volume is computationally intensive. If volume is chosen as the estimation performance index, then the optimization problem in (\ref{E8-20}) lacks an analytical solution and requires significantly more computing resources to solve.
	\end{remark}
	
	\begin{lemma}
		\label{L8-3}
		With the parameter matrices determined in Lemma \ref{L8-2}, the DZFE $\hat{\mathcal{X}}_{[1]}^{\mathrm{opt}}(k)$ will provide better estimation performance compared to any original local zonotopic estimates, i.e.,
		\begin{equation}
			\label{E8-24}
			\left\| R_{[1]}^{\mathrm{opt}}(k) \right\|_W^2 \le \| R_{i}(k) \|_W^2 \ \ (i=1,2,...,L)
		\end{equation}
		where $R_{[1]}^{\mathrm{opt}}(k):=\min_{M(k)} \left\| R_{[1]}(k) \right\|_W^2$ can be obtained by substituting (\ref{E8-21}) into (\ref{E8-18}).
	\end{lemma}
	
	\noindent \textbf{Proof.}
		The proof of this Lemma is straightforward. If $M_i(k)=\mathbf{I} \ (i \in \mathbb{N}_L \setminus \{1\})$ and $M_j(k)=\mathbf{0} \ (j \in \mathbb{N}_L \setminus \{1,i\})$, then $\left\|R_{[1]}(k)\right\|_W^2=\| R_{i}(k) \|_W^2$. If $M_i(k)=\mathbf{0} \ (i \in \mathbb{N}_L \setminus \{1\})$, then $\left\|R_{[1]}(k)\right\|_W^2=\| R_{1}(k) \|_W^2$. Notice that $R_{[1]}^{\mathrm{opt}}(k)$ with the parameter matrices obtained in (\ref{E8-20}) are optimal; thus (\ref{E8-24}) holds.
		\qed
		\vspace{8pt}
	
	Although the DZFE with optimal parameters in Lemma \ref{L8-2} can provide better estimation performance than original local zonotopic estimates, the fusion result is still conservative (see Section V for details). In what follows, we will provide a more accurate DZFE inspired by our previous work \cite{zhang2023distributed}. Firstly, the $\mathcal{H}$-representation of $\hat{\mathcal{X}}_{[1]}^{\mathrm{opt}}(k)$ can be written as
	\begin{equation}
		\label{E8-25}
		\hat{\mathcal{X}}_{[1]}^{\mathrm{opt}}(k) = \left\{ x(k) \in \mathbb{R}^n: H(k) x(k) \le b(k) \right\}
	\end{equation}
	where $H(k) \in \mathbb{R}^{2\varepsilon \times n}$ and $b(k) \in \mathbb{R}^{2\varepsilon}$ are denoted by
	\begin{equation}
		\label{E8-57}
		\begin{cases}
			& \!\!\!\!\!\!
			H(k) = \mathrm{col} \{ h_1(k), -h_1(k), \cdots, h_{\varepsilon}(k), -h_{\varepsilon}(k) \}
			\\
			& \!\!\!\!\!\!
			b(k) = \mathrm{col} \{b_1(k), \cdots, b_{2\varepsilon}(k)\}.
		\end{cases}
	\end{equation}
	 Obviously, the intersection of local zonotopic estimates $\mathcal{O}(k)$ is still a polytope. Let us denote the $\mathcal{V}$-representation of $\mathcal{O}(k)$ as:
	\begin{equation}
		\label{E8-28}
		\mathcal{O}(k) = \mathrm{cone}(V(k))
	\end{equation}
	where $V(k)=\begin{bmatrix} \nu_1(k) & \cdots & \nu_{\eta}(k) \end{bmatrix}$. Notice that a pair of parallel hyperplanes from $\hat{\mathcal{X}}_{[1]}^{\mathrm{opt}}(k)$, i.e., $\{x(k): h_i(k) x(k) = b_{2i-1}(k)\}$ and $\{x(k): h_i(k) x(k) = -b_{2i}(k)\}$ may not constitute a pair of tight parallel hyperplanes for the intersection $\mathcal{O}(k)$. It is proposed to reduce the distance between the parallel hyperplanes to make them a pair of tight parallel hyperplanes.
	
	Given the center $\hat x_{[1]}^{\mathrm{opt}}(k)$, one needs to find strips with the normal vector $h_i(k) \ (i \in \mathbb{N}_{\varepsilon})$ that exactly enclose $\mathcal{O}(k)$. The first hyperplane is the one that contains a vertex of $\mathcal{O}(k)$ and has the maximum distance to the center. It can be searched by enumeration from the following problem:
	\begin{equation}
		\label{E8-29}
		\mathbf{s}_i(k)= \arg \max_{j \in \mathbb{N}_{\eta}} \left|h_i(k) \left( \hat x_{[1]}^{\mathrm{opt}}(k) - \nu_j(k) \right)\right|.
	\end{equation}
	Then, the two parallel hyperplanes can be represented by
	\begin{subequations}
		\label{E8-30}
		\begin{align}
			& \mathcal{P}_{i,1}(k) \!:=\! \left\{ z(k)  \in  \mathbb{R}^n : h_i(k)z(k)  \! \le \! h_i(k) \nu_{ \mathbf{s}_i(k) }(k) \right\}
			\\
			& 
			\begin{aligned}
				\mathcal{P}_{i,2}(k) \!:=
				&\Big\{ z(k) \in \mathbb{R}^n : h_i(k)z(k)  \le  2 h_i(k) \hat x_{[1]}^{\mathrm{opt}}(k)
				\\
				& \ \ \ \ \ \ \ \ \ \ \ \ \ \ \ \ \ \ \ \ \ \ \ \ \ \ \ - h_i(k) \nu_{ \mathbf{s}_i(k) }(k) \Big\}.
			\end{aligned}
		\end{align}
	\end{subequations}
	It is known that parallel hyperplanes (\ref{E8-30}) can be transformed into a strip $\mathcal{S}(c_i(k),d_i(k)) \ (i \in \mathbb{N}_{\varepsilon})$. Then, the following theorem is proposed to construct the improved version of zonotope fusion criterion.
	
	\begin{theorem}
		\label{T8-2}
		Given the DZFE with optimal parameters $\hat{\mathcal{X}}_{[1]}^{\mathrm{opt}}(k)$ as in Lemma \ref{L8-3}, an improved DZFE $\hat{\mathcal{X}}_{[2]}(k)$ can be determined by the following fusion criterion:
		\begin{equation}
			\label{E8-31}
			\hat{\mathcal{X}}_{[2]}(k) = \bigcap\nolimits_{i=1}^{\mathbb{N}_{\varepsilon}} \mathcal{S}(c_i(k),d_i(k)).
		\end{equation}
		Moreover, the improved DZFE obtained by (\ref{E8-31}) satisfies $x(k) \in \mathcal{O}(k) \subseteq \hat{\mathcal{X}}_{[2]}(k) \subseteq \hat{\mathcal{X}}_{[1]}(k)$.
	\end{theorem}
	
	\noindent \textbf{Proof.}
		By the construction of strips $\mathcal{S}(c_i(k),d_i(k)) \ (i=1,...,\varepsilon)$ in (\ref{E8-29}) and (\ref{E8-30}), one has that
		\begin{equation}
			\label{E8-33}
			\mathcal{O}(k) \subseteq \bigcap\nolimits_{i=1}^{\mathbb{N}_{\varepsilon}} \mathcal{S}(c_i(k),d_i(k)).
		\end{equation}
		Therefore, it can be inferred that $x(k) \in \mathcal{O}(k) \subseteq \hat{\mathcal{X}}_{[2]}(k)$. As the parallel hyperplanes in (\ref{E8-30}) are tight parallel hyperplanes for the intersection $\mathcal{O}(k)$, it follows that $\hat{\mathcal{X}}_{[2]}(k) \subseteq \hat{\mathcal{X}}_{[1]}^{\mathrm{opt}}(k)$. 
		\qed
		\vspace{8pt}
	
	\begin{lemma}
		\label{L8-4}
		The improved DZFE $\hat{\mathcal{X}}_{[2]}(k)$ in Theorem \ref{T8-2} can achieve better estimation performance than any original local zonotopic estimates, i.e.,
		\begin{equation}
			\label{E8-34}
			\left\| R_{[2]}(k) \right\|_W^2 \le \| R_{i}(k) \|_W^2 \ \ (i=1,2,...,L).
		\end{equation}
	\end{lemma}
	
	\noindent \textbf{Proof.}
		The proof of this lemma follows by Lemma \ref{L8-3} and the fact that $\hat{\mathcal{X}}_{[2]}(k) \subseteq \hat{\mathcal{X}}_{[1]}^{\mathrm{opt}}(k)$.
		\qed
		\vspace{8pt}
	
	Then, the computation procedures for the improved version of distributed zonotopic fusion estimator are summarized in Algorithm 1.
	
	\label{Algo:1}
	\begin{algorithm}[h]
		\caption{Improved version of distributed zonotopic fusion estimation.}
		\begin{algorithmic}[1]
			\STATE \textbf{Initialization}: Given $\hat{\mathcal{X}}_i(0) \ (i \in \mathbb{N}_L)$ for each subsystem such that $x(0) \in \hat{\mathcal{X}}_i(0)$;
			\FOR{$k = 1,2, ...$}
			\FOR{$i = 1,2,...,L$}
			\STATE Subsystem $\mathbf{S}_i$ collects its local measurement $y_i(k)$ and calculates its local zonotopic estimate $\hat{\mathcal{X}}_i(k)$ as in (\ref{E8-52}-\ref{E8-54});
			\STATE Subsystem $\mathbf{S}_i$ transmits its local zonotopic estimate $\hat{\mathcal{X}}_i(k)$ to the FC.
			\ENDFOR
			\STATE The FC solves the optimization problem (\ref{E8-20}) with the analytical solution (\ref{E8-21}) to obtain the optimal parameter matrices $M(k)$;
			\STATE The FC computes the DZFE with optimal parameters $\hat{\mathcal{X}}_{[1]}^{\mathrm{opt}}(k)$ by (\ref{E8-15});
			\STATE The FC computes the $\mathcal{H}$-representation of $\hat{\mathcal{X}}_{[1]}^{\mathrm{opt}}(k)$ as in (\ref{E8-25}) and the $\mathcal{V}$-representation of the intersection of local zonotopic estimates $\mathcal{O}(k)$ as in (\ref{E8-28});
			\STATE The FC solves the problem (\ref{E8-29}) by enumeration to find out tight strips $\mathcal{S}(c_i(k),d_i(k))$ for $\mathcal{O}(k)$;
			\STATE The FC computes the improved DZFE $\hat{\mathcal{X}}_{[2]}(k)$ by (\ref{E8-31}).
			\ENDFOR	
		\end{algorithmic}
	\end{algorithm}
	
	\begin{remark}
		\label{R8-8}
		Notice that the improved DZFE also satisfies the state inclusion property because $\mathcal{O}(k) \subseteq \hat{\mathcal{X}}_{[2]}(k)$. It means that $\hat{\mathcal{X}}_{[2]}(k)$ can inherit good properties from $\hat{\mathcal{X}}_{[1]}(k)$, including the performance superiority in Lemma \ref{L8-4} and the state inclusion property. Compared with the fusion criterion in our previous work \cite{zhang2023distributed} using a parallelotope enclosure, the proposed fusion criterion in Theorem \ref{T8-2} has additional performance superiority, which is more reliable and useful for zonotope fusion. Moreover, the fusion criterion in Theorem 2 is less conservative because the general zonotopes can describe more accurate sets than parallelotopes.
	\end{remark}
	
	\subsection{Sequential zonotope fusion criterion}
\begin{figure}[h]
		\begin{center}
			\includegraphics[height=2.48cm, width=8.8cm]{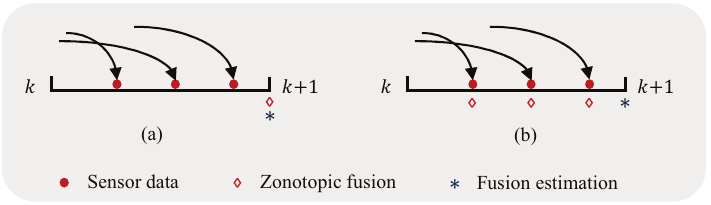}
			\caption{The diagram of two fusion strategies: (a) batch zonotope fusion; (b) sequential zonotope fusion.}
			\label{fig8-4} 
		\end{center}
	\end{figure}

	The proposed zonotope fusion criteria in Theorems \ref{T8-1}-\ref{T8-2} are of batch fusion strategies. Only when local zonotopic estimates are all available at the fusion center, the fusion step can be implemented to fusion them all at a fusion instant. However, the specific time for the fusion center to complete data receiving for different sensors is different during the time period from instant $k-1$ to instant $k$. It means that such batch fusion strategies may induce long computation delay \cite{zhang2018sequential} and are not appropriate for real-time applications. Therefore, a sequential zonotope fusion criterion need to be designed by combining the local zonotopic estimates one by one according to the time order of data receiving. The diagram of batch and sequential zonotope fusion strategies are plotted in Fig. \ref{fig8-4}.
	
	Without loss of generality, suppose that $L$ local zonotopic estimates obtained in time order as $\hat{\mathcal{X}}_{\vec{1}}(k)$, ..., $\hat{\mathcal{X}}_{\vec{i}}(k)$, ..., $\hat{\mathcal{X}}_{\vec{L}}(k)$, where $\hat{\mathcal{X}}_{\vec{i}}(k):= \langle \hat x_{\vec{i}}(k), R_{\vec{i}}(k) \rangle$ is the $i$th arrived local zonotopic estimate which satisfies $\hat{\mathcal{X}}_{\vec{i}}(k) \in \left\{\hat{\mathcal{X}}_{1}(k), \hat{\mathcal{X}}_{2}(k), ..., \hat{\mathcal{X}}_{L}(k)\right\}$. The sequential zonotope fusion criterion contains $L$ fusion stages and one improvement stage. The initial fusion stage is $\hat{\mathcal{X}}_{f_1}(k)=\hat{\mathcal{X}}_{\vec{1}}(k)$, and the other fusion stages are represented as
	\begin{equation}
		\label{E8-35}
		\hat{\mathcal{X}}_{f_i}(k) = f\left( \hat{\mathcal{X}}_{f_{i-1}}(k), \hat{\mathcal{X}}_{\vec{i}}(k) \right) \ \ (i =2,3,...,L)
	\end{equation} 
	where $\hat{\mathcal{X}}_{f_i}(k):=\langle \hat x_{f_i}(k), R_{f_i}(k) \rangle \ (i \in \mathbb{N}_L)$ is the $i$th fused result. The sequential DZFE is then obtained after using the improvement strategy to $\hat{\mathcal{X}}_{f_L}(k)$ as in Theorem 2. The function $f(\cdot)$ is a sequential zonotope fusion criterion and can be designed in the following theorem.
	
	\begin{theorem}
		\label{T8-3}
		Given the local zonotopic estimates in time order $\hat{\mathcal{X}}_{\vec{i}}(k) \ (i=2,...L)$, the $i$th sequential fusion stage in (\ref{E8-35}) yields $\hat{\mathcal{X}}_{f_i}(k)$ as follows:
		\begin{equation}
			\label{E8-36}
			\begin{cases}
				& \!\!\!\!\!\!
				\hat x_{f_i}(k) = \hat x_{f_{i-1}}(k) + M_i(k) \left( \hat x_{\vec{i}}(k) - \hat x_{f_{i-1}}(k) \right)
				\\
				& \!\!\!\!\!\!
				R_{f_i}(k) = \big[ \left( I\!-\! M_i(k) \right)\!R_{f_{i-1}}(k) \ \ M_i(k)R_{\vec{i}}(k) \big]
			\end{cases}
		\end{equation}
		where $M_i(k) \in \mathbb{R}^{n \times n}$ is a designed parameter matrix. The optimal parameter matrices $M_i^{\mathrm{opt}}(k) \ (i=1,2...,L)$ which enable the fusion results in (\ref{E8-36}) to achieve the optimal estimation performance (\ref{E8-14}) at every stages can be determined by the following optimization problem:
		\begin{equation}
			\label{E8-37}
			\begin{aligned}
				M_i^{\mathrm{opt}}(k) 
				& = \arg \min_{M_i(k)}  J \left( R_{f_i}(k) \right)
				\\
				&= \arg  \min_{M_i(k)}  \left\| R_{f_i}(k) \right\|_W^2.
			\end{aligned}
		\end{equation}
		Moreover, the analytical solution of (\ref{E8-37}) is found to be
		\begin{equation}
			\label{E8-51}
			M_i^{\mathrm{opt}}(k) \!=\! R_{f_{i\!-\!1}}\!(k) \left( R_{f_{i\!-\!1}}\!(k) R_{f_{i\!-\!1}}^{\mathrm{T}}\!(k) \!+\! R_{\vec{i}}(k) R_{\vec{i}}^{\mathrm{T}}(k) \right)^{\!-1}\!\!.
		\end{equation}
	\end{theorem}
	
	\noindent \textbf{Proof.}
		Similar to the proof of Lemma \ref{L8-2}, the derivative of the optimization objective (\ref{E8-37}) with respect to $M_i(k)$ is as follows:
		\begin{equation}
			\label{E8-38}
			\!
			\begin{aligned}
				\frac{\partial \left\| R_{f_i}(k) \right\|_W^2}{\partial M_i(k)} 
				& = \frac{\partial \left\| \left( I- M_i(k) \right) R_{f_{i-1}}(k) \right\|_W^2}{\partial M_i(k)} 
				\\
				& \ \ \ \ +  \frac{\partial \left\| M_i(k)R_{\vec{i}}(k) \right\|_W^2}{\partial M_i(k)}.
			\end{aligned}
		\end{equation}
		By the definition of weighted Frobenius norm and the properties of trace derivatives, one has that
		\begin{equation}
			\label{E8-39}
			\!
			\begin{aligned}
				\frac{\partial \left\| R_{f_i}(k) \right\|_W^2}{\partial M_i(k)} 
				& = 2W M_i(k) R_{f_{i-1}}(k) R_{f_{i-1}}^{\mathrm{T}}(k)
				\\
				& \!\!\!\! - 2W R_{f_{i-1}}(k) + 2W M_i(k) R_{\vec{i}}(k) R_{\vec{i}}^{\mathrm{T}}(k).
			\end{aligned}
		\end{equation}
		It is known from Lemma \ref{L8-1} and (\ref{E8-36}) that $\left\| R_{f_i}(k) \right\|_W^2$ is convex with respect to $M_i(k)$. Therefore, the optimal estimation performance is achieved when $\partial \left\| R_{f_i}(k) \right\|_W^2/\partial M_i(k)=0$, and the analytical solution in (\ref{E8-51}) follows from (\ref{E8-39}).
		\qed
		\vspace{8pt}
	
	The sequential DZFE obtained from Theorem \ref{T8-3} with the improvement strategy also satisfies the state inclusion property and performance superiority. We will prove them in the following lemmas.
	
	\begin{lemma}
		\label{L8-5}
		The sequential DZFE in (\ref{E8-35}-\ref{E8-36}) satisfies state inclusion property, i.e., $x(k) \in \mathcal{O}(k) \subseteq \hat{\mathcal{X}}_{[3]}(k)$
	\end{lemma}
	
	\noindent \textbf{Proof.}
		First of all, we will prove that $\hat{\mathcal{X}}_{f_{i-1}}(k) \bigcap \hat{\mathcal{X}}_{\vec{i}}(k) \subseteq \hat{\mathcal{X}}_{f_{i}}(k)$ for the $i$the fusion stage in Theorem \ref{T8-3}. By the general-representation of a zonotope, if $x(k) \in \hat{\mathcal{X}}_{f_{i-1}}(k) \bigcap \hat{\mathcal{X}}_{\vec{i}}(k)$, then there exist $u_{f_{i-1}}(k) \in \mathcal{B}^{r(i-1)}$ and $u_{\vec{i}}(k) \in \mathcal{B}^r$ such that
		\begin{subequations}
			\label{E8-40}
			\begin{align}
				& x(k) - \hat x_{f_{i-1}}(k) - R_{f_{i-1}}(k) u_{f_{i-1}}(k) = 0
				\\
				& \hat x_{f_{i-1}}(k) \!+\! R_{f_{i-1}}(k) u_{f_{i-1}}(k) \!- \hat x_{\vec{i}}(k) \!- R_{\vec{i}}(k) u_{\vec{i}}(k) \!=\! 0.
			\end{align}
		\end{subequations}
		The equations in (\ref{E8-40}) is actually a sufficient condition for the following equation:
		\begin{equation}
			\label{E8-41}
			x(k)  = \hat x_{f_i}(k) + R_{f_i}(k) u_{f_i}(k)
		\end{equation}
		where $u_{f_i}(k):=\mathrm{col}\{u_{f_{i-1}}(k),u_{\vec{i}}(k)\} \in \mathcal{B}^{ri}$. It means that $x(k) \in \hat{\mathcal{X}}_{f_i}(k)$ by the general-representation of a zonotope, and $\hat{\mathcal{X}}_{f_{i-1}}(k) \bigcap \hat{\mathcal{X}}_{\vec{i}}(k) \subseteq \hat{\mathcal{X}}_{f_i}(k)$ can be inferred from (\ref{E8-40}) to (\ref{E8-41}). Then, we will prove this lemma by induction. It is known from the initial fusion stage that $x(k) \in \hat{\mathcal{X}}_{\vec{1}}(k) = \hat{\mathcal{X}}_{f_1}(k)$. For the $i$th fusion stage in (\ref{E8-36}), assume that $x(k) \in \bigcap_{j=1}^{i-1} \hat{\mathcal{X}}_{\vec{j}}(k) \subseteq \hat{\mathcal{X}}_{f_{i-1}}(k)$, then $x(k) \in \bigcap_{j=1}^{i} \hat{\mathcal{X}}_{\vec{j}}(k) \subseteq \hat{\mathcal{X}}_{f_{i-1}}(k) \bigcap \hat{\mathcal{X}}_{\vec{i}}(k) \subseteq \hat{\mathcal{X}}_{f_{i}}(k)$ holds by previous derivation. By induction, one can conclude that $x(k) \in \bigcap_{i=1}^{L} \hat{\mathcal{X}}_{\vec{i}}(k) \subseteq \hat{\mathcal{X}}_{f_L}(k)$. By Theorem \ref{T8-2}, we know that the improved version of zonotope fusion can keep the property that $\hat{\mathcal{X}}_{[3]}(k) \subseteq \hat{\mathcal{X}}_{f_L}(k)$, which means that $x(k) \in \mathcal{O}(k) \subseteq \hat{\mathcal{X}}_{[3]}(k)$.
		\qed
		\vspace{8pt}
	
	\begin{lemma}
		\label{L8-6}
		The sequential DZFE in (\ref{E8-35}-\ref{E8-36}) with optimal parameters in (\ref{E8-37}) can achieve better estimation performance than any original local zonotopic estimates, i.e.,
		\begin{equation}
			\label{E8-42}
			\left\| R_{[3]}(k) \right\|_W^2 \le \| R_{i}(k) \|_W^2 \ \ (i=1,2,...,L).
		\end{equation}
	\end{lemma}
	
	\noindent \textbf{Proof.}
		The proof is similar to the derivation of Lemmas \ref{L8-3}-\ref{L8-4}; thus omitted.
		\qed
		\vspace{8pt}
	
	\begin{remark}
		\label{R8-9}
		The sequential DZFE can avoid long computation delay by fusing local zonotopic estimates one by one. At the same time, the overall calculation time can also be reduced because the high-dimensional matrix inverse operation in (\ref{E8-21}) is simplified into $L-1$ low-dimensional matrix inverse operations in (\ref{E8-51}).
	\end{remark}
	
	The computation procedures for the sequential zonotopic fusion estimator are summarized in Algorithm 2.
	
	\label{Algo:2}
	\begin{algorithm}[h]
		\caption{Sequential zonotopic fusion estimation.}
		\begin{algorithmic}[1]
			\STATE \textbf{Initialization}: Given $\hat{\mathcal{X}}_i(0) \ (i \in \mathbb{N}_L)$ for each subsystem such that $x(0) \in \hat{\mathcal{X}}_i(0)$;
			\FOR{$k = 1,2, ...$}
			\FOR{$i = 1,2,...,L$}
			\STATE Subsystem $\mathbf{S}_i$ collects its local measurement $y_i(k)$ and calculates its local zonotopic estimate $\hat{\mathcal{X}}_i(k)$ as in (\ref{E8-52}-\ref{E8-54});
			\STATE Subsystem $\mathbf{S}_i$ transmits its local zonotopic estimate $\hat{\mathcal{X}}_i(k)$ to the FC.
			\ENDFOR
			\STATE The FC initializes the fusion stage $\hat{\mathcal{X}}_{f_1}(k)=\hat{\mathcal{X}}_{\vec{1}}(k)$;
			\FOR{$i=2,...,L$}
			\STATE The FC solves the optimization problem (\ref{E8-37}) with the analytical solution (\ref{E8-51}) to obtain the optimal parameter matrices $M_i(k)$;
			\STATE The FC computes the $i$th fusion result $\hat{\mathcal{X}}_{f_i}(k)$ by (\ref{E8-36});
			\ENDFOR
			\STATE The FC obtains the sequential DZFE $\hat{\mathcal{X}}_{[3]}(k)=\hat{\mathcal{X}}_{f_L}(k)$ by using the improvement strategy in Theorem 2.
			\ENDFOR	
		\end{algorithmic}
	\end{algorithm}
	
	\subsection{Stability Analysis}
	To analyze the stability conditions for the proposed DZFEs, we will introduce some lemmas. It is known from (\ref{E8-54}) that $R_i(k) = \downarrow_{r,W} \left( R_i^o(k) \right)$. The following lemma from \cite{combastel2015zonotopes} shows the performance relationship before and after the weighted reduction operation.
	
	\begin{lemma}
		\label{L8-7}
		Given a spd matrix $W_i$ with all its eigenvalues in $[\underline{\lambda}_i, \overline{\lambda}_i] \subset \mathbb{R}, \underline{\lambda}_i > 0$, the performance relationship between $R_i^o(k) \in \mathbb{R}^{n \times \bar r}$ and $R_i(k) \in \mathbb{R}^{n \times r}$ satisfies
		\begin{equation}
			\label{E8-43}
			\left\| R_i(k) \right\|_{W_i}^2 \le \left( 1+\frac{\mu_i}{d+r} \right) \left\| R_i^o(k) \right\|_{W_i}^2
		\end{equation}
		where $d=\bar r-r$ and $\mu_i = ((\overline{\lambda}_i/\underline{\lambda}_i)(d+n)-1)(d+n)$.
	\end{lemma}
	
	The following lemma from \cite{combastel2015zonotopes} provides conditions to keep a linear time-varying system robustly $\gamma$-stable.
	
	\begin{lemma}
		\label{L8-8}
		The time-varying discrete-time linear system $x(k) = A(k-1) x(k-1)$ is robustly $\gamma$-stable (i.e., stable with decay rate $\gamma$) if there exists a bounded sequence of spd matrices $W(k)$ and $\gamma \in [0,1]$ such that, $\forall k \ge 0$,
		\begin{equation}
			\label{E8-44}
			\begin{bmatrix}
				\gamma W(k-1) & A^{\mathrm{T}}(k-1) W(k) \\
				\star & W(k)
			\end{bmatrix} \succ 0.
		\end{equation}
	\end{lemma}
	
	\begin{assumption}
		\label{A8-1}
		A bounded sequence of local estimator gain matrices $K_i^{*}(k)$ and a bounded sequence of spd matrices $W_i^{*}(k)$ have been designed such that $x_i(k) = (\mathbf{I}-K_i^{*}(k)C_i(k))A(k-1) x_i(k-1)$ is robustly $\gamma_i$-stable, where $\gamma_i <1$ and $W_i^*(k) \ (\forall k >0)$ have all their eigenvalues within a constant real interval $[\underline{\lambda}_i, \overline{\lambda}_i], \underline{\lambda}_i > 0$.
	\end{assumption}
	
	The following lemma benefits from the performance superiority of the proposed DZFEs.
	
	\begin{lemma}
		\label{L8-9}
		The proposed DZFEs in Theorems \ref{T8-1}-\ref{T8-3} are ultimately bounded if one of the local zonotopic estimates is ultimately bounded, i.e.,
		\begin{equation}
			\label{E8-45}
			\begin{aligned}
				& \lim_{k \rightarrow \infty} \! \| R_i(k) \|_{W_i^*(k)}^2 < p_i \ (\exists i \in \mathbb{N}_L)  \\
				& \ \ \ \ \ \ \ \ \ \ \ \ \ \ \ \ \ \ \ \ \ \ \ \Longrightarrow \ \lim_{k \rightarrow \infty} \! \| R(k) \|_{W_i^*(k)}^2 < p_i
			\end{aligned}
		\end{equation}
		where $p_i$ is a positive finite number.
	\end{lemma}
	
	\noindent \textbf{Proof.}
		By the performance superiority in Lemmas \ref{L8-3}, \ref{L8-4}, and \ref{L8-6}, one has that $\| R(k) \|_W^2 \le \| R_i(k) \|_W^2 \ (i \in \mathbb{N}_L)$, then this lemma can be obtained directly.
		\qed
		\vspace{8pt}
	
	\begin{theorem}
		\label{T8-4}
		The proposed DZFEs in Theorems \ref{T8-1}-\ref{T8-3} are ultimately bounded if there exists $i \in \mathbb{N}_L$ such that $\gamma_i \left( \! 1\!+\!\frac{\mu_i}{d+r} \! \right) < 1$ holds.
	\end{theorem}
	
	\noindent \textbf{Proof.}
		According to Assumption 1 and (\ref{E8-44}) in Lemma \ref{L8-8}, one has the following inequality from Schur complement lemma \cite{boyd1994linear}:
		\begin{equation}
			\label{E8-46}
			\begin{aligned}
				\gamma_i W_i^*(k-1) \succ 
				& A^{\mathrm{T}}(k-1) (\mathbf{I}-K_i^{*}(k)C_i(k))^{\mathrm{T}} W_i^{*}(k)
				\\
				& \times (\mathbf{I}-K_i^{*}(k)C_i(k)) A(k-1).
			\end{aligned}
		\end{equation}
		After left multiplication by $R_i^{\mathrm{T}}(k-1)$ and right multiplication by $R_i(k-1)$ to both sides of (\ref{E8-46}), the following inequality can be derived by the fact that the trace of a negative definite matrix is negative:
		\begin{equation}
			\label{E8-47}
			\begin{aligned}
				& \| (\mathbf{I}-K_i^{*}(k)C_i(k))A(k-1) R_i(k-1) \|_{W_i^{*}(k)}^2
				\\
				& \ \ \ \ \ \ \ \ \ \ \ \ \ \ \ \ \ \ \ \ \ \ \ \ \ \ \ \ \ \ \ \ < \gamma_i \|R_i(k-1)\|_{W_i^*(k-1)}^2.
			\end{aligned}
		\end{equation}
		Since the optimal local estimator gain is used in each sensor, one can conclude by (\ref{E8-43}) and (\ref{E8-47}) that
		\begin{equation}
			\label{E8-48}
			\begin{aligned}
				\left\| R_i(k) \right\|_{W_i^*(k)}^2 < 
				& \gamma_i \left( 1+\frac{\mu_i}{d+r} \right)  \|R_i(k-1)\|_{W_i^*(k-1)}^2 
				\\
				& + \phi_i(k)
			\end{aligned}
		\end{equation}
		where $\phi_i(k):=\| (\mathbf{I}-K_i(k)C_i(k))A(k-1) B(k-1) \|_{W_i^{*}(k)}^2 + \|K_i(k)D_i(k)\|_{W_i^{*}(k)}^2$ is a finite constant term due to the boundedness of system matrices. Therefore, if $\gamma_i \left( \! 1\!+\!\frac{\mu_i}{d+r} \! \right) < 1$ holds, then
		\begin{equation}
			\label{E8-49}
			\lim_{k \rightarrow \infty} \left\| R_i(k) \right\|_{W_i^*(k)}^2 < p_i.
		\end{equation}
		By Lemma \ref{L8-9}, this theorem is then proven.
		\qed
		\vspace{8pt}
	
	\section{Numerical Examples}
	In this section, we demonstrate the advantages of the proposed zonotope fusion criteria by fusing 2-dimensional zonotopes and evaluate the effectiveness of the proposed DZFEs using a target tracking example.
	
	\subsection{Fusion of Two-dimensional Zonotopes}
	\begin{figure}[h]
		\begin{center}
			\includegraphics[height=13cm, width=8.5cm]{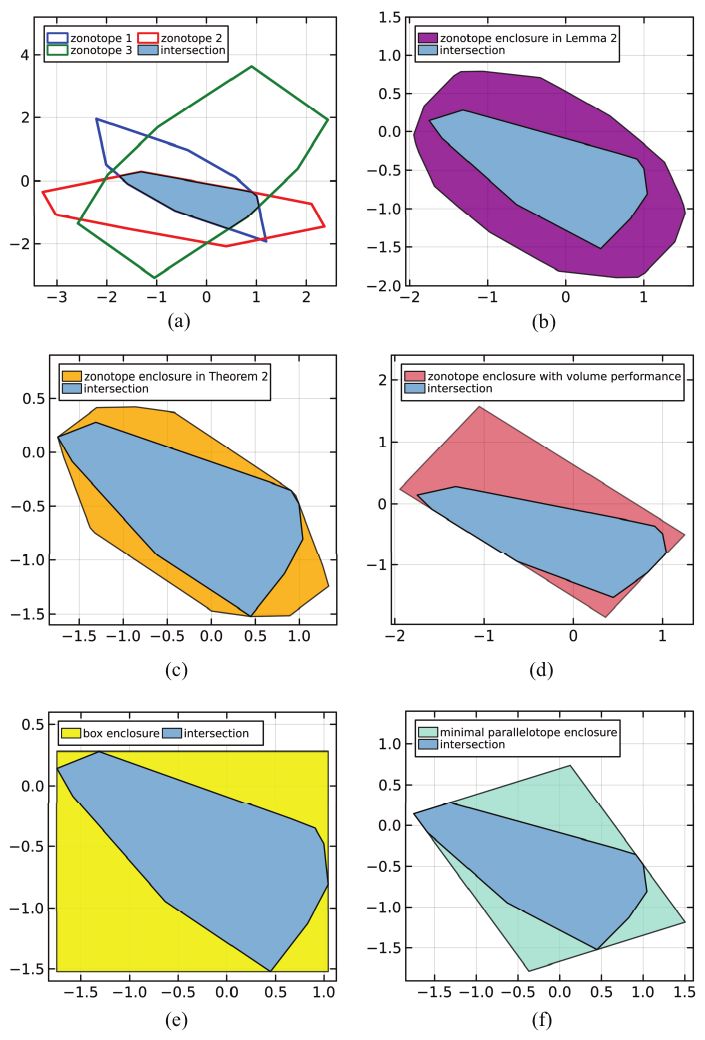}
			\caption{(a) Three zonotopes and their intersection. (b) The minimal zonotope enclosure by Lemma 2. (c) The improved zonotope enclosure by Theorem 2. (d) The minimal zonotope enclosure under volume performance. (e)The box enclosure. (f) The minimum parallelotope enclosure.}  
			\label{fig8-2}
		\end{center}
	\end{figure}
	\begin{figure}[h]
		\begin{center}
			\includegraphics[height=7cm, width=8.5cm]{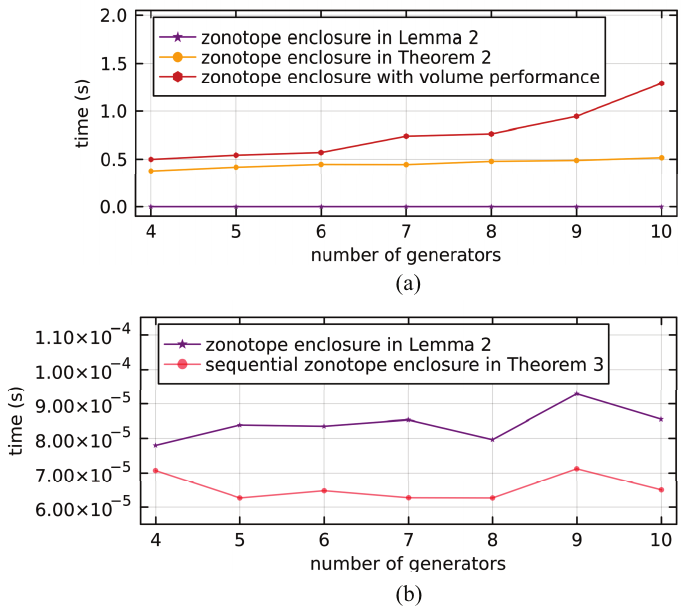}
			\caption{The computing time of zonotope fusion criteria.}
			\label{fig8-7}
		\end{center}
	\end{figure}
	In this example, three 2-dimensional 4-order zonotopes are fused using different zonotope fusion criteria. The original zonotopes and their intersection are plotted in Fig. \ref{fig8-2} (a). The minimum zonotope enclosure in (\ref{E8-15}) under the performance index (\ref{E8-14}) is plotted in Fig. \ref{fig8-2} (b), which is computed by the analytical solution in Lemma \ref{L8-2}. To reduce conservatism, an improved zonotope enclosure is computed by Theorem \ref{T8-2} and plotted in Fig. \ref{fig8-2} (c). It can be seen from Fig. \ref{fig8-2} (b-c) that the improvement strategy from Theorem \ref{T8-2} significantly reduces the size of the zonotope enclosure. Additionally, the minimum zonotope enclosure in (\ref{E8-15}) under the volume performance index is obtained by solving the first optimization problem in (\ref{E8-20}) and is presented in Fig. \ref{fig8-2} (d). It is noteworthy that the nonlinear optimization problem in (\ref{E8-20}) is computationally inefficient and does not offer obvious performance advantages compared to the improved zonotope fusion in Theorem \ref{T8-2}. Furthermore, the box enclosure and the minimum parallelotope enclosure from \cite{zhang2023distributed} are plotted in Fig. \ref{fig8-2} (e-f). As discussed in Remark \ref{R8-8}, these two enclosures do not provide performance superiority over the original zonotopes. To evaluate computational efficiency, 2-dimensional zonotopes with different numbers of generators are randomly generated for zonotope fusion. The computing time for zonotope fusion criteria has been plotted in Fig. \ref{fig8-7}. As shown in Fig. \ref{fig8-7} (a), the minimum zonotope enclosure under the volume performance index needs more time to solve optimization problems and its computing time increases exponentially with the number of generators, whereas the other two fusion criteria exhibit a slower increase. To illustrate the advantages of sequential zonotope fusion, the total computing time for fusion stages is compared to batch zonotope fusion in Lemma \ref{L8-2}. The results, shown in Fig. \ref{fig8-7} (b), indicate that sequential zonotope fusion requires less computational time as it only involves low-dimensional matrix computation.

	\subsection{Target Tracking with Multi-sensors}
	\begin{figure}[h]
		\begin{center}
			\includegraphics[height=8cm, width=8.5cm]{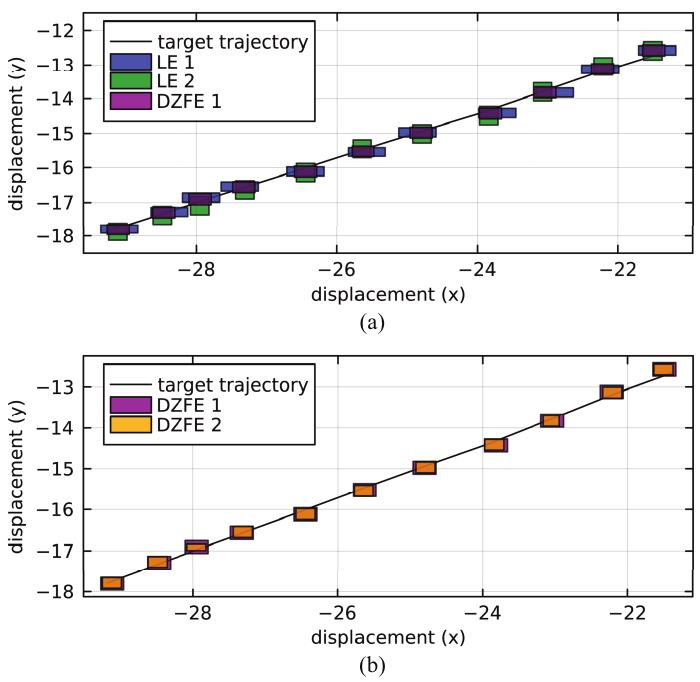}
			\caption{The trajectory of the moving target and the projections of zonotopic estimates in $(d_{\mathbf{x}}, d_{\mathbf{y}})$: (a) two local zonotopic estimates and the DZFE with optimal parameters; (b) the DZFE with optimal parameters and the improved DZFE.}  
			\label{fig8-3}
		\end{center}
	\end{figure}
	\begin{figure}[h]
		\begin{center}
			\includegraphics[height=8cm, width=8.5cm]{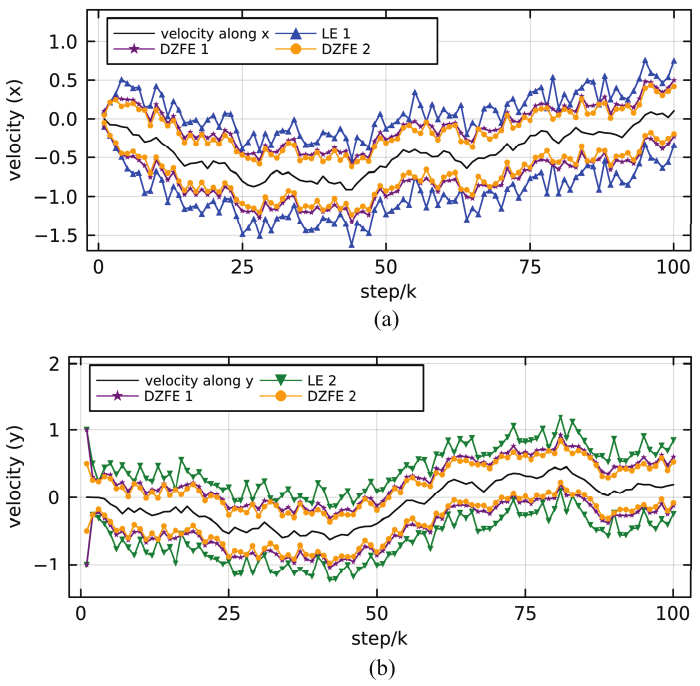}
			\caption{The velocity of the moving target and the projections of the local zonotopic estimates, the DZFE with optimal parameters, and the improved DZFE: (a) along $\mathbf{x}$ direction; (b) along $\mathbf{y}$ direction.}
			\label{fig8-5} 
		\end{center}
	\end{figure}
	\begin{figure}[h]
		\begin{center}
			\includegraphics[height=6.08cm, width=6.5cm]{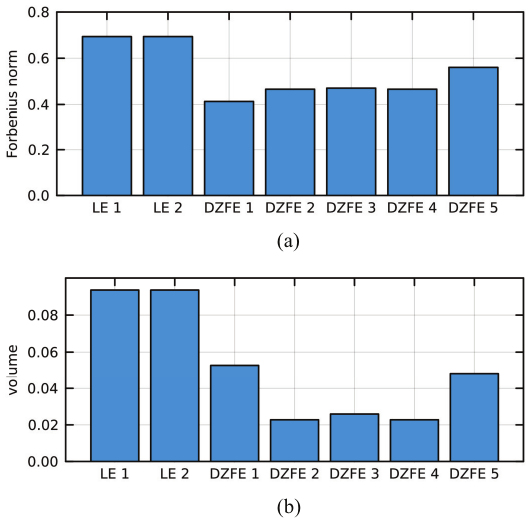}
			\caption{The estimation performance evaluated by: (a) the weighted Frobenius norm of zonotopes; (b) the volume of zonotopes.}  
			\label{fig8-6}
		\end{center}
	\end{figure}
	In the second example, we consider a target tracking system with two sensors. The dynamics of a non-maneuvering target is described as follows \cite{rongli2003survey}:
	\begin{equation}
		\label{E8-50}
		x(k+1) \!=\! \begin{bmatrix} 1 & T & 0 & 0 \\ 0 & 1 & 0 & 0 \\ 0 & 0 & 1 & T \\ 0 & 0 & 0 & 1 \end{bmatrix} \!x(k) + \begin{bmatrix} T^2/2 & 0 \\ T & 0 \\ 0 & T^2/2 \\ 0 & T \end{bmatrix}\!w(k)
	\end{equation}
	where $x:=\left[ d_\mathbf{x} \ v_\mathbf{x} \ d_\mathbf{y} \ v_\mathbf{y} \right]^{\mathrm{T}}$ represents the target's state, and $w(k)$ denotes the system's process noise. The parameter $T$ is the sampling period, while $(d_\mathbf{x}, d_\mathbf{y})$ and $(v_\mathbf{x}, v_\mathbf{y})$ are the displacement and velocity coordinates along the $\mathbf{x}$ and the $\mathbf{y}$ axes, respectively. The displacement coordinates of the non-maneuvering target are observed by two sensors with different precision, with the measurement model represented as follows:
	\begin{equation}
		\label{E8-56}
		\begin{cases}
			& \!\!\!\!\!\!
			y_1(k) = \begin{bmatrix} 1 & 0 & 0 & 0 \\ 0 & 0 & 1 & 0 \end{bmatrix} x(k) + \begin{bmatrix} 2 & 0 \\ 0 & 1 \end{bmatrix} v_1(k)
			\\
			& \!\!\!\!\!\!
			y_2(k) = \begin{bmatrix} 0 & 0 & 1 & 0 \\ 1 & 0 & 0 & 0 \end{bmatrix} x(k) + \begin{bmatrix} 1 & 0 \\ 0 & 2 \end{bmatrix} v_2(k).
		\end{cases}
	\end{equation}
	Here, $y_i(k) \ (i=1, 2)$ and $v_i(k) \ (i=1, 2) $ denote the measurements and the measurement noises, respectively.

	In the simulation, the sampling period is set to $T = 1$, and both process noises and measurement noises are bounded within the zonotope $\langle \textbf{0}, \textbf{I} \rangle$. Two local zonotopic estimators are deployed in sensors to estimate the state of the moving target. These local estimates are then transmitted to a fusion center via a communication network for zonotope fusion. Different zonotopic fusion estimates are computed based on different zonotope fusion criteria. DZFE 1 is obtained by computing the analytical solution provided in Lemma \ref{L8-2}, which is optimal under the performance index (\ref{E8-14}). Meanwhile, DZFE 2 is computed from Theorem \ref{T8-2} using the improvement strategy. To show the effectiveness of the proposed fusion estimators, the computed zonotopes are projected onto the displacement coordinates $(d_{\mathbf{x}}, d_{\mathbf{y}})$. In Fig. \ref{fig8-3}, the target trajectory and the projections of zonotopic estimates during time steps $40$ to $50$ are depicted. It is evident that the projection of DZFE 1 in $(d_{\mathbf{x}}, d_{\mathbf{y}})$ is significantly smaller than that of the local zonotopic estimates, whereas the projection of DZFE 2 is fully wrapped within that of DZFE 1. Both of these two fusion estimators outperform the local estimators in terms of estimation performance, with DZFE 2 being less conservative than DZFE 1. Moreover, the computed zonotopes are projected onto the coordinate $v_{\mathbf{x}}$ and the coordinate $v_{\mathbf{y}}$, as shown in Fig. \ref{fig8-5}. It can be observed that the projected interval of DZFE 1 along $\mathbf{x}$ direction is within that of the first local zonotopic estimate, while the projected interval of DZFE 1 along $\mathbf{y}$ direction is within that of the second local zonotopic estimate. This is attributed to the first sensor having better precision in the in $\mathbf{y}$ direction and the second sensor having better precision in $\mathbf{x}$ direction. Consistent with the improvement strategy, the projected intervals of DZFE 2 fall within those of DZFE 1. In addition, DZFEs 3-5 are computed by the optimization problem in (\ref{E8-20}) under volume performance index, the box enclosure, and the minimal parallelotope enclosure, respectively. The weighted Frobenius norms and volumes of these DZFEs are calculated to better demonstrate their performance, with the average values over time plotted in Fig. \ref{fig8-6}. In this figure, LE 1 and LE 2 represent the two local zonotopic estimates. The results indicate that all fusion estimates outperform local estimates, regardless of weighted Frobenius norms and volumes. Particularly, DZFE 1 achieves the minimal weighted Frobenius norm, while DZFE 2 achieves the minimal volume, thereby illustrating the effectiveness of the improvement strategy.
	
	\section{Conclusion}
	In this paper, we investigated the problem of distributed zonotopic fusion estimation. Three zonotope fusion criteria were designed to integrate local zonotopic estimates from sensor nodes for enhancing estimation performance. The designed DZFEs outperform local zonotopic estimates in terms of estimation performance and preserve the state inclusion property. The first DZFE can achieve optimal performance through an optimization problem, with its analytic solution provided in Lemma \ref{L8-2}. The second DZFE improves the zonotope enclosure by constructing tight strips for the intersection of local zonotopic estimates. The third DZFE employs a sequential fusing strategy based on the arrival order of local zonotopic estimates, which reduces computational burden and communication delay. Their stability conditions were derived based on the performance superiority, which indicates that the boundedness of the fusion estimates depends on the existence of at least one bounded local estimate. The effectiveness of the proposed methods has been demonstrated through illustrative examples.
	
	Multi-sensor fusion estimation is a field that integrates theoretical and practical aspects deeply. Future work will focus on applying the proposed zonotope fusion criteria to more practical scenarios, such as cooperative localization and path planning.
	
	\bibliographystyle{ieeetr}
	\bibliography{08}
	
	\begin{IEEEbiography}[{\includegraphics[width=1in,height=1.23in,clip,keepaspectratio]{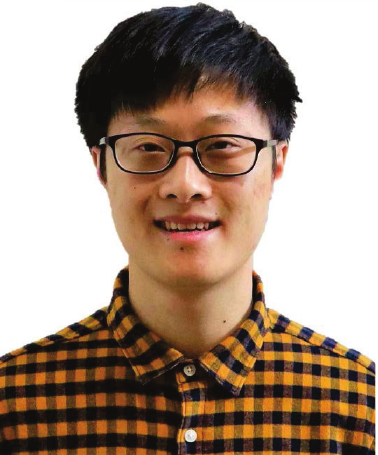}}]{Yuchen Zhang} (Member, IEEE) received the B.Eng. degree in Electrical Engineering and Automation and the Ph.D. degree in Control Theory and Control Engineering from Zhejiang University of Technology, Hangzhou, China, in 2017 and 2024, respectively.
		
		His current research interests include multi-sensor fusion estimation, distributed estimation and control of interconnected systems, privacy-preserving estimation, and set-based estimation and optimization. 
	\end{IEEEbiography}
	
	\begin{IEEEbiography}[{\includegraphics[width=1in,height=1.25in,clip,keepaspectratio]{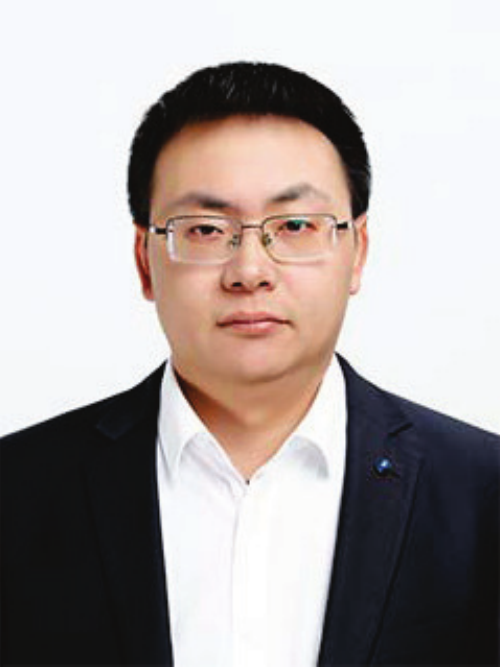}}]{Bo Chen} (Senior Member, IEEE) received the B.S. degree in information and computing science from Jiangxi University of Science and Technology, Ganzhou, China, in 2008, and the Ph.D degree in Control Theory and Control Engineering from Zhejiang University of Technology, Hangzhou, China, in 2014.
		
		He jointed the Department of Automation, Zhejiang University of Technology in 2018, where he is currently a Professor. He was a Research Fellow with the School of Electrical and Electronic Engineering, Nanyang Technological University, Singapore, from 2014 to 2015 and from 2017 to 2018. He was also a Postdoctoral Research Fellow with the Department of Mathematics, City University of Hong Kong, Hong Kong, from 2015 to 2017. His current research interests  include information fusion, distributed estimation and control, networked fusion systems, and secure estimation of cyber-physical systems.
		
		Prof. Chen was a recipient of the Outstanding Thesis Award of Chinese Association of Automation in 2015 and also was a recipient of the First Prize of Natural Science of Ministry of Education in 2020. He serves as Associate Editor for IET Control Theory and Applications, Journal of the Franklin Institute, and Frontiers in Control Engineering, and also serves as Guest Editor for IEEE Transactions on Industrial Cyber-Physical Systems.
	\end{IEEEbiography}

	\begin{IEEEbiography}[{\includegraphics[width=1in,height=1.25in,clip,keepaspectratio]{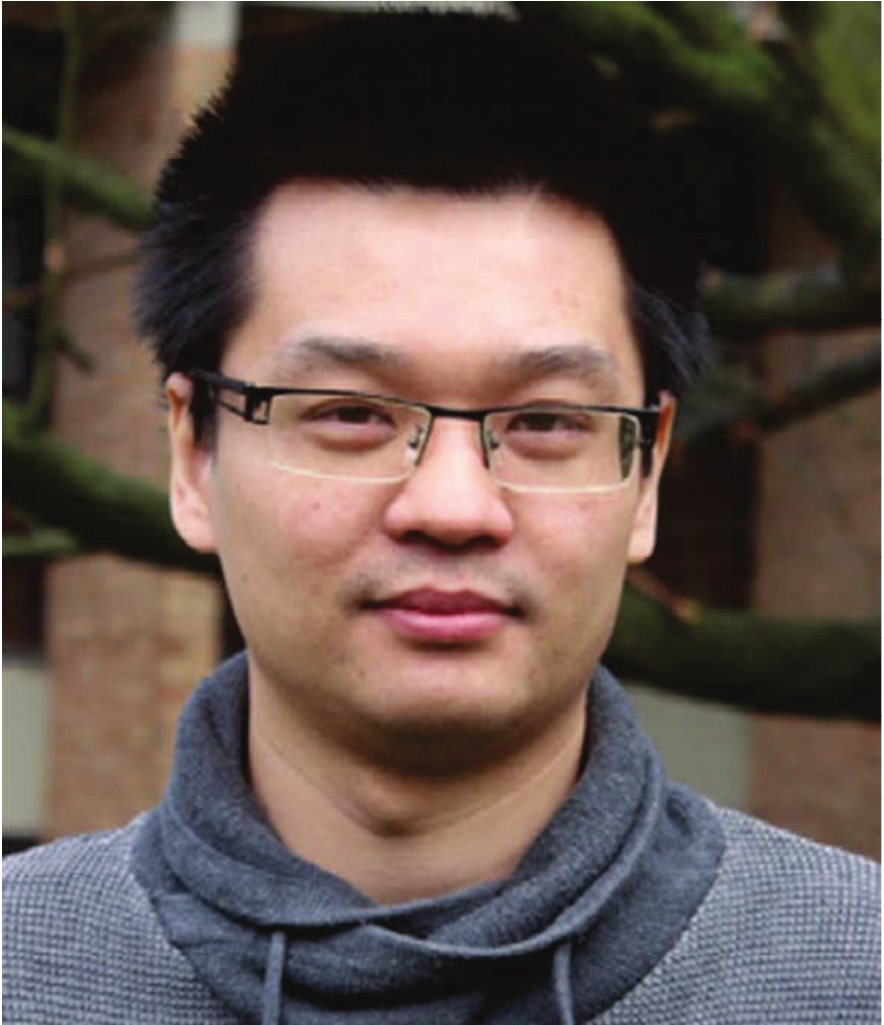}}]{Zheming Wang} (Member, IEEE) received his B.S. degree in Mechanical Engineering \& Automation (Mechatronics) from the Department of Mechanical Engineering of Shanghai Jiao Tong University in 2012 and Ph.D. degree in Control and Mechatronics from the Department of Mechanical Engineering of National University of Singapore in 2016. He is currently a Professor in the Department of Automation at Zhejiang University of Technology. His main research interests lie in model predictive control, set-theoretic methods and data-driven control. He serves as an Associate Editor for IEEE CSS Conference Editorial Board, the IEEE CSS Technology Conference Editorial Board, and the IFAC/Elsevier journal Nonlinear Analysis: Hybrid Systems.
	\end{IEEEbiography}

	\begin{IEEEbiography}[{\includegraphics[width=1in,height=1.25in,clip,keepaspectratio]{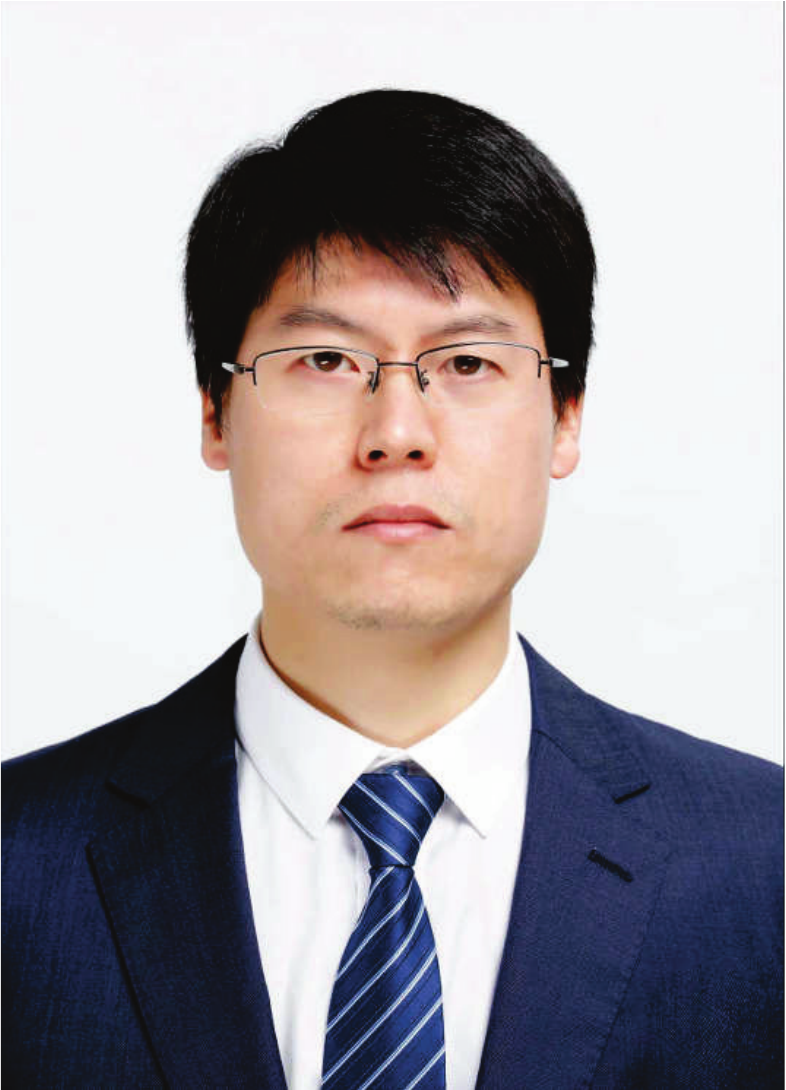}}]{Wen-An Zhang} (Senior Member, IEEE) received the B.Eng. degree in Automation and the Ph.D. degree in Control Theory and Control Engineering from Zhejiang University of Technology, China, in 2004 and 2010, respectively. He has been with Zhejiang University of Technology since 2010 where he is now a professor at Department of Automation. He was a senior research associate at Department of Manufacturing Engineering and Engineering Management, City University of Hong Kong, 2010-2011. He was awarded an Alexander von Humboldt Fellowship in 2011-2012. His current research interests include multi-sensor information fusion estimation and its applications. He has been serving as a subject editor for Optimal Control Applications and Methods from September	2016.
	\end{IEEEbiography}
	
	\begin{IEEEbiography}[{\includegraphics[width=1in,height=1.25in,clip,keepaspectratio]{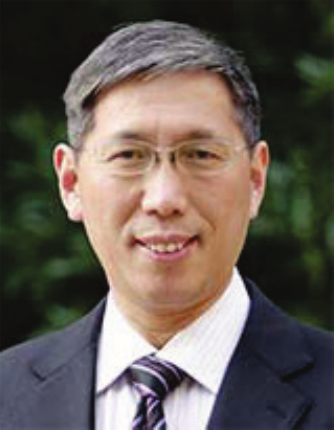}}]{Li Yu} (Senior Member, IEEE) received the B.S. degree in control theory from Nankai University, Tianjin, China, in 1982, and the M.S. and Ph.D. degrees from Zhejiang University, Hangzhou, China, in 1988 and 1999, respectively.
		
		He is currently a Professor with the College of Information Engineering, Zhejiang University of Technology, Hangzhou, China. He has successively presided over 20 research projects. He has published five academic monographs, one textbook, and over 300 journal papers. He has also been authorized over 100 patents for invention and granted five scientific and technological awards. His current research interests include robust control, networked
		control systems, cyber–physical systems security, and information fusion.
	\end{IEEEbiography}

	\begin{IEEEbiography}[{\includegraphics[width=1in,height=1.25in,clip,keepaspectratio]{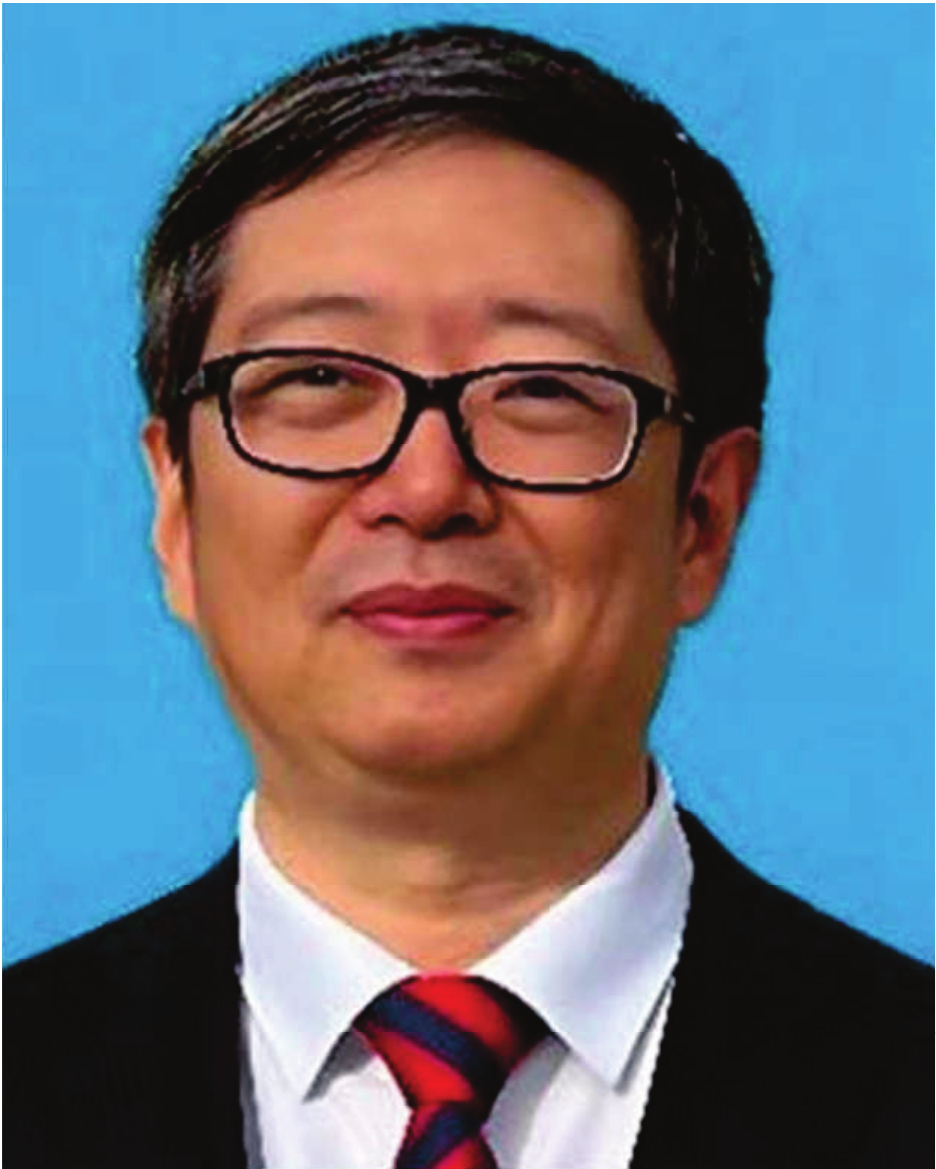}}]{Lei Guo} (Fellow, IEEE) was born in Qufu, China, in 1966. He received the B.S. degree in mathematics and M.S. degree in operations research and cybernetics from Qufu Normal University, Qufu, in 1988 and 1991, respectively, and the Ph.D. degree in automatic control and applications from Southeast University, Nanjing, China, in 1997.
		
		From 1991 to 1994, he was a Lecturer with Qingdao University, Qingdao, China. He has published more than 480 papers, seven monographs, and has more than 180 authorized invention patents. His research interests include anti-disturbance control theory and applications, intelligent navigation and control technology of unmanned systems.
		
		Prof. Guo was aPostdoctoral Fellow and Associate Fellow with South east University, from 1997 to 1999. From 1999 to 2000, he was a Postdoctoral Fellow with IRCCyN, Nantes, Frances. From 2000 to 2005, he was a Research Associate/Fellow/Visiting Professor with Loughborough University, UMIST and Manchester University, U.K. From 2004 to 2006, he was a Professor with Southeast University, now he is a Guest Chair Professor. From 2006 to now, he is a Distinguish Professor and Director of the Space Intelligent Control Research Center, Beihang University, Beijing, China. He is an Academician of the Chinese Academy of Sciences (CAS), a Fellow of IEEE, IET, Chinese Association of Automation (CAA), and China Association of Inventions (CAI). He is the Director of the Navigation, Guidance and Control Committee of the CAA. He was the recipient of the National Nature Science Awards (2013), National Technology Invention Awards (2018), National Pioneer Innovation Award (2023) of China. He also obtained the Gold Medal of International Exhibition of Inventions of Geneva, Nuremberg and Turkey for bio-inspired navigation sensor, compound-eye-inspired navigation systems and biomimetic flying robots, respectively.
	\end{IEEEbiography}

\end{document}